\documentclass[aps, prl, twocolumn, floatfix, superscriptaddress, reprint, 10pt]{revtex4-1}  
\usepackage{bm}
\usepackage{graphicx}
\usepackage[usenames]{color}
\usepackage{microtype}
\usepackage{amsmath}
\usepackage{amssymb}
\usepackage{xspace}
\usepackage{comment}
\usepackage{footnote}
\usepackage{relsize}
\usepackage{hhline}
\usepackage{hyperref}

\usepackage{physics}

\newcommand{\br}{\boldsymbol{r}}
\newcommand{\be}{\begin{equation}}
\newcommand{\ee}{\end{equation}}
\usepackage{floatrow}
\newfloatcommand{capbtabbox}{table}[][\FBwidth]

\newcommand{\editor}[2]{%
  \expandafter\newcommand\csname #1note\endcsname[1]{%
    \textcolor{#2}{(\textbf{#1:} ##1)}}%
  \expandafter\newcommand\csname #1\endcsname[1]{%
    \textcolor{#2}{##1}}%
  \expandafter\newcommand\csname #1cancel\endcsname[1]{%
    \textcolor{#2}{\sout{##1}}}%
  \expandafter\newcommand\csname #1change\endcsname[2]{%
    \textcolor{#2}{\sout{##1} ##2}}%
  \newenvironment{#1text}{\color{#2}}{\color{black}}
}

\begin{document}

\title{
Predicting the Charge Density Response in Metal Electrodes 
}

\author{Andrea Grisafi}
\email{andrea.grisafi@ens.psl.eu}
\affiliation{PASTEUR, D\'epartement de chimie, \'Ecole Normale Sup\'erieure, PSL University, Sorbonne Universit\'e, CNRS, 75005 Paris, France}

\author{Augustin Bussy}
\affiliation{Department of Chemistry, University of Zurich, Winterthurerstrasse 190, 8057 Z\"urich, Switzerland}

\author{{Mathieu Salanne}}
\affiliation{ 
{Sorbonne Universit\'e, CNRS, Physicochimie des \'Electrolytes et Nanosyst\`emes Interfaciaux, F-75005 Paris, France}
}
\affiliation{{Institut Universitaire de France (IUF), 75231 Paris, France}}

\author{Rodolphe Vuilleumier}
\affiliation{PASTEUR, D\'epartement de chimie, \'Ecole Normale Sup\'erieure, PSL University, Sorbonne Universit\'e, CNRS, 75005 Paris, France}

\begin{abstract}
The computational study of energy storage and conversion processes calls for simulation techniques that can reproduce the electronic response of metal electrodes under electric fields. Despite recent advancements in machine-learning  methods applied to electronic-structure properties, predicting the non-local behavior of the charge density in electronic conductors remains a major open challenge. We~combine long-range and equivariant kernel methods to predict the Kohn-Sham electron density of metal electrodes {in response to various kinds of electric field perturbations}. By taking slabs of gold as an example, we first show how the non-local electronic polarization generated by the interaction with an ionic species can be accurately reproduced in electrodes of arbitrary thickness. A finite-field extension of the method is then introduced, which allows us to predict the charge transfer and the electrostatic potential drop induced by the application of a homogeneous and constant electric field. {Finally, we demonstrate the capability of the method to reproduce the charge-density response in a gold/electrolyte capacitor under an applied voltage, predicting the system polarization with a greater accuracy than state-of-the-art classical atomic-charge models.} 
\end{abstract}

\maketitle

The behaviour of the electronic charge density in metal surfaces plays a decisive role in the study of energy storage and conversion processes occurring in batteries, capacitors and electrocatalytic frameworks~\cite{Lautar2020jacs,Beinlich2022,Karmodak2022}. Its accurate computer simulation ultimately requires the adoption of first-principles methods that are capable to predict the non-local response of an electronic conductor subject to an external perturbation. Density functional theory (DFT), in particular, has been widely applied to the description of metal interfaces in various research areas of surface chemistry and catalysis~\cite{norskov2011,Grosjean2019,Gerrits2020}. Depending on the specific application, different methods can be adopted to keep the metal at a constant charge~\cite{Souza2013,Freysoldt2020}, fixed applied potential~\cite{Lozovoi2001,Filhol2006,Letchworth2012,Hoermann2019,Melander2019,Dominguez2023}, as well as to induce a macroscopic polarization in the system due to the presence of an applied field~\cite{Stengel2007,Che2017,Ke2022}. In spite of the success of these approaches, the extensive use of DFT to the study of realistic metal interfaces is however hindered by an unfavorable scaling with the system size, effectively limiting their applicability to a few hundreds of atoms~\cite{Khatib2021,Chen2022}.

Beyond first-principles approaches, classical molecular dynamics (MD) methods have also gained major attention in the context of enabling the simulation of metal interfaces at the nanometric scale~\cite{Jeanmairet2022}. These methods go from the adoption of fluctuating atomic charges~\cite{Siepmann1995,Reed2007,Reed2008,Limmer2013,Scalfi2020} to the use of image-charge boundary conditions~\cite{Petersen2012,Dwelle2019}. 
The downside of MD is of course the lack of an explicit quantum treatment, which limits the description of the electrode either to a perfect metal approximation, or to the adoption of a finite screening length computed from semiclassical Thomas-Fermi models~\cite{Scalfi2020thomas-fermi,Scalfi2021}. For these reasons, reaching the level of accuracy of DFT in the large-scale simulation of the metal charge density represents a goal of paramount importance~\cite{DiPasquale2021,DiPasquale2023}.

In recent years, many efforts have been devoted to the development of machine-learning (ML) methods dedicated to the prediction of the electronic charge density~\cite{broc+17nc,alre+18cst,chan+19npjcm,gris+19acscs,fabr+19cs,bogo+20nc,Jorgensen2021,lewis+21jctc,Grisafi2023,Rackers2023}. The success of these methods is mostly grounded on the interplay between a local decomposition of the scalar field and the adoption of local representations of the atomic structure that are used as input vectors of the ML model. 
When it comes to conducting systems, however, local ML models are expected to show strong limitations in predicting the variations of the charge density over large distances, especially in the presence of external fields. 
In fact, while some ML methods that explicitly incorporate long-range effects have already been developed~\cite{Grisafi2019jcp,Gao2022,Zhang-Weinan2022,Westermayr2022}, 
the study of charge-transfer phenomena in electronic conductors has to date been limited to simplified charge-equilibration schemes that represent the electron density via a set of atomic charges~\cite{Ghasemi2015,Ko2021,Staacke2022}. Moreover, an explicit ML-treatment of the charge density in metal surfaces under electric fields has thus far only been investigated by virtue of suitable response functions that enter conceptual DFT approaches~\cite{Shao2022,Dufils2023arxiv}. In this letter, we show how to combine long-range and equivariant learning methods to accurately predict the charge density response in metal electrodes, {including the derivation of a finite-field model suitable to treat the application of external electric fields.} 

Let us start by considering the linear expansion of the electron density on an atom-centered basis $\chi_{n\lambda\mu}$ given by the product of radial Gaussian-type functions $R^\lambda_n$ and spherical harmonics~$Y_{\lambda\mu}$:
\begin{equation}\label{eq:density-fitting}
    n_e(\br) \approx \sum_{in\lambda\mu} \, c_i^{n\lambda\mu} \sum_{ix,iy} \chi_{n\lambda\mu}(\br-\boldsymbol{r}_i-\boldsymbol{u}_{ix,iy})\, ,
\end{equation}
where $\boldsymbol{r}_i$ are the atomic positions in the unit cell and $\boldsymbol{u}_{ix,iy}$ are the cell translation vectors used to account for the two dimensional periodicity of the metallic surface. We assume that the coefficients $c_i^{n\lambda\mu}$ come from a density-fitting procedure of a reference Kohn-Sham density.

The decomposition of Eq.~\eqref{eq:density-fitting} can be used within equivariant ML models to predict $n_e$ in a  highly transferable (atom-centered) fashion~\cite{gris+19acscs,Rackers2023}. In this work, we rely on a recently optimized kernel-based method
introduced in Ref.~\cite{Grisafi2023}, also known as SALTED. Within SALTED, a linear approximation of the density expansion coefficients is provided which satisfies the rotational symmetry of spherical harmonics, i.e., $\boldsymbol{c}_i^{n\lambda} = \sum_{M}\boldsymbol{k}_{iM}^{\lambda}\, \boldsymbol{b}_M^{n\lambda}$. Here, $\boldsymbol{b}_M^{n\lambda}$ are vector-valued regression weights of dimension $(2\lambda+1)$ and $\boldsymbol{k}_{iM}^{\lambda}$ is a symmetry-adapted kernel matrix of dimension $(2\lambda+1)^{\otimes2}$ which encodes the similarity between the local structural features of atom $i$ and those of a sparse selection of atoms $\{M\}$ belonging to the training set~\cite{gris+18prl}. The complexity of the learning problem then comes down to the adoption of physically inspired structural representations of the atomic environment~\cite{musil_chem_rev}, $\boldsymbol{P}_i^{\lambda}$, which define the kernel matrix as an inner product over a suitable feature space~\cite{deringer_chem_rev}, i.e., $\boldsymbol{k}_{iM}^{\lambda} = \boldsymbol{P}_{i}^{\lambda} \boldsymbol{P}_M^{\lambda^\dag}$. 

A typical choice for $\boldsymbol{P}_i^{\lambda}$ consists in an equivariant generalization of the popular smooth overlap of atomic positions (SOAP) method~\cite{bart+13prb,gris+18prl}. This approach ultimately derives from the definition of a smooth Gaussian-density distribution describing the local environment of atom $i$, i.e., $\rho_i(\br) = f_\text{cut}(r)\sum_j e^{-\alpha|\br-\br_{ij}|^2} $, with $j$ running over the atomic neighbours and  $f_\text{cut}(r)$ a spherical cutoff function of radius~$r_\text{cut}$. 
By construction, the model thus neglects any long-range effect that occurs beyond~$r_\text{cut}$ in exchange of a high level of transferability. While this is generally not a problem~\cite{fabr+20chimia,lewis+21jctc}, we aim to demonstrate that endowing the structural features with a long-range character is essential when dealing with metal surfaces. For this reason, we rely on a definition of $\boldsymbol{P}_i^{\lambda}$ based on long-distance equivariant (LODE) representations~\cite{gris-ceri19jcp}. In particular, we adopt an implementation of LODE that combines information about the local atomic density $\rho_i(\br)$ and a Hartree-like potential originated by the Gaussian density of the atoms of the entire system~\cite{grisafi2021cs}: 
\begin{equation}\label{eq:lode-potential}
    V_i(\br) = f_\text{cut}(r)\, \int d\br' \frac{\sum_j e^{-\alpha|\br'-\br_{ij}|^2}}{|\br'-\br|} \, .
\end{equation}
Note that the cutoff function is here applied \textit{after} the Coulomb operator, thus guaranteeing the inclusion of long-range information within the local environment of $i$. The descriptor of order $\lambda$ is finally obtained from a symmetry-adapted tensor product of $\rho_i$ and $V_i$ expanded on a set of orthogonal radial and angular functions~\cite{willatt2019,grisafi2021cs}: 
\begin{equation}\label{eq:lode-descriptor}
  P_i^{\lambda\mu}(nn'll') = \sum_{mm'} \rho^{nlm}_i\, V^{n'l'm'}_i\braket{lm,l'm'}{\lambda\mu}\, ,
\end{equation}
where $n$ and $lm$ are the radial and angular indexes, respectively, and where $\braket{lm,l'm'}{\lambda\mu}$ are the Clebsch-Gordan coefficients used for the composition of angular momenta.

\begin{figure}[t!]
    \centering
    \includegraphics[width=8.5cm]{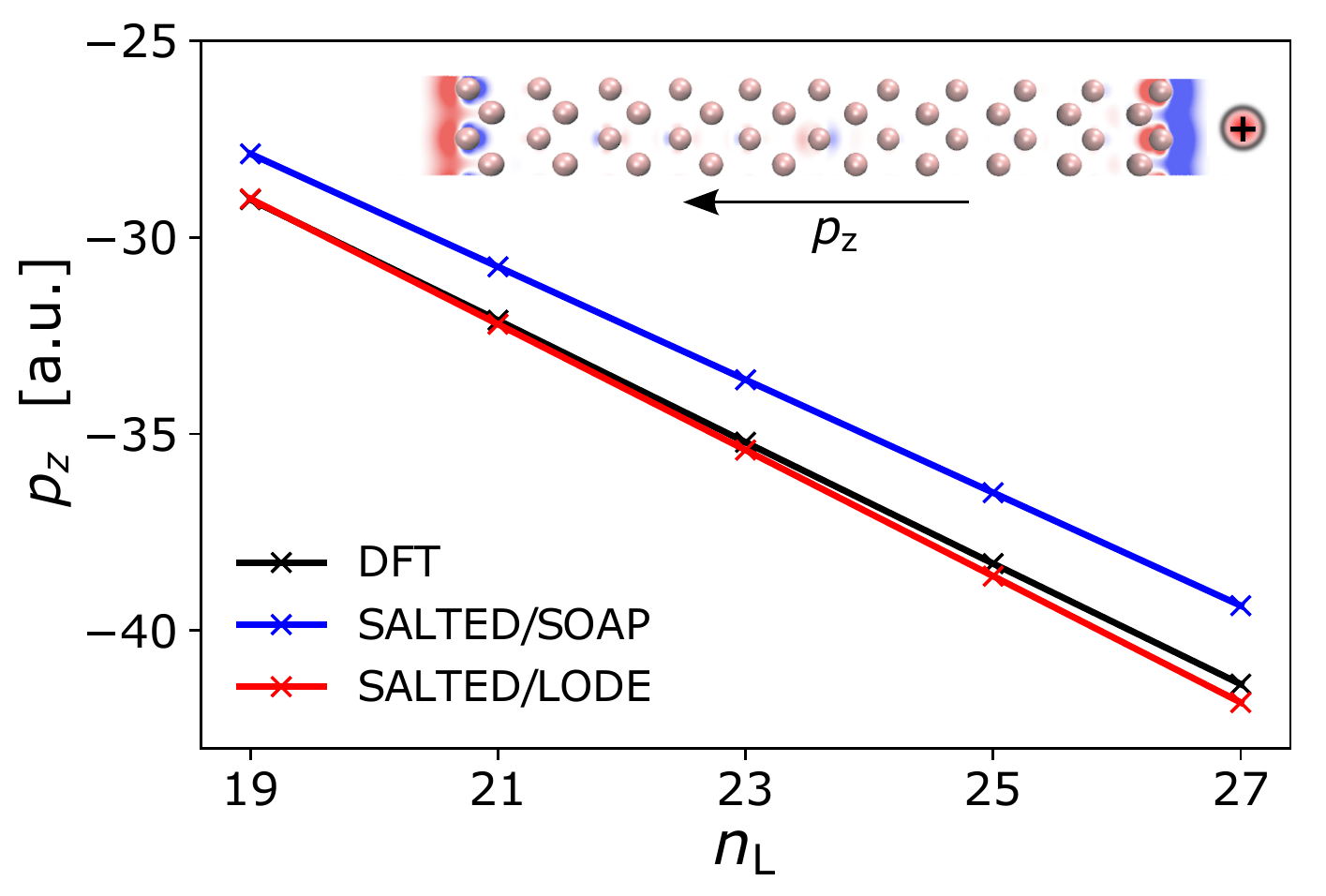}
    \caption{{Extrapolated electronic polarization $p_z$ of Au(100) electrodes of increasing thickness $n_\text{L}$, induced by a Na$^+$ cation placed at 4{\AA} distance from the upper metal surface. Predictions are obtained for SALTED/SOAP (blue line) and SALTED/LODE (red line) models trained on the electron-density response of electrodes that include up to $n_\text{L}$=15 atomic layers. Black line: reference DFT polarization. Inset: volume slice of the predicted charge density response for a $n_\text{L}$=21 gold electrode; color code from blue to red corresponds to a linear scale from $-$0.001 to +0.001 a.u., respectively.}} 
    \label{fig:fig1}
\end{figure}

We start by considering slabs of gold aligned perpendicularly to the $z$-axis, which interact with a sodium cation Na$^+$ placed at 4~{\AA} distance from the upper metal surface. In particular, we consider symmetric Au(100) slabs made of 2 unit cell repetitions along the $xy$ plane and spanning from 3 to 15 gold layers along~$z$. A total of 240 training configurations are then generated by taking uniform random displacements of the atomic positions up to 2.5\% of the lattice constant along the three Cartesian directions. The electrode polarization is simulated at the QM/MM level of theory by representing the ion as a classical Gaussian charge~\cite{Laino2006}. 
Reference density coefficients~$c^{n\lambda\mu}_i$ for the gold electrodes are obtained by performing calculations of $n_e$ at the DFT/PBE level~\cite{Perdew1996}, in combination with a density-fitting approach based on an overlap metric~\cite{Vahtras1993,briling2021,Bussy2023}. {The learning target is finally defined as the difference between the perturbed electron densities and the electron densities of the corresponding isolated electrodes, i.e., $\Delta n_e = n_e-n_e^0$.} 

From a ML point of view, the ion-induced polarization of the metal electrode is expected to be naturally captured by a multispecies treatment of Eq.~\eqref{eq:lode-descriptor}. This accounts for computing as many $\rho_i$ and~$V_i$ as the number of  chemical species~$a$ in the system~\cite{will+18pccp,grisafi2021cs}, {while letting $i$ run exclusively on the gold atoms. Upon this procedure, we train SALTED/LODE models using a local cutoff of $r_\text{cut}$~=~8{\AA}, where the atom density and potential fields are defined from Gaussian widths of $\sigma=0.5${\AA} and $\sigma=4.0${\AA}, respectively. For comparison, SALTED/SOAP models are similarly trained by substituting $V_i$ with $\rho_i$ in Eq.~\eqref{eq:lode-descriptor}}. Further details about the reference calculations and machine-learning parameters are reported in the Supplemental Material (SM),~Ref.~\cite{suppmat}.

We put the method to the test on rigid gold electrodes that include from $n_\text{L}$=19 to $n_\text{L}$=27 atomic layers. In so doing, we extrapolate the charge-density response over a range of distances between the two metal surfaces that extends well beyond that spanned by the training set. As an accuracy measure of the ion-induced charge transfer, we compute the metal electronic polarization  along $z$, $p_z$, which can be analytically obtained from the predicted density coefficients~\cite{suppmat}. Fig.~\ref{fig:fig1} reports the prediction results as a function of the number of atomic layers~$n_\text{L}$. {As expected, we find that a local SOAP-based model is unable to reproduce the long-range character of the charge-density response, resulting in predicted polarization values that are off by  $\sim$36\% of the standard deviation of $p_z$ in the test set.} Conversely, SALTED/LODE predictions are found to accurately extrapolate the electronic polarization at increasing electrode thicknesses, thus capturing the expected linear decrease of $p_z$ with respect to $n_\text{L}$. This is confirmed by the density-derived calculation of the Hartree potential drop through the metallic slab, which is found in very good agreement with that of DFT for all the test electrodes considered~\cite{suppmat}. {We note that performing these predictions took $\sim$1 second per structure on a single node with 24 CPUs, resulting in a speedup of the order of $3 \times10^3$ with respect to DFT~\cite{suppmat}.}

Having shown the importance of long-range structural information in describing the response properties of metal electrodes, we now proceed with investigating the charge transfer effect induced by an applied electric field. From here on, we will only refer to SALTED/LODE models. The accumulation/depletion of electronic charge at the two metal surfaces is simulated by performing  calculations  under a constant and uniform electric field along~$z$. The field intensity is chosen as $E_z=-1.0$~V/{\AA}, which, for a perfect metal, corresponds to an induced surface charge of $\sigma=\pm $5.53$\times 10^{-3}$~e/{\AA}$^2$. DFT densities are generated from the same gold configurations and using the same level of theory already adopted in the previous example. 

In order to reproduce the polarization of the electrode induced by an applied electric field~$E_z$, the axial symmetry of the system around~$z$ must be explicitly incorporated into the ML model. To tackle this problem, 
we start by defining a local external potential field centered at the position of a given atom $i$:
\begin{equation}
    V^{E}_i(\boldsymbol{r}) = E_z\,(z-z_i)f_\text{cut}(r)\, .
\end{equation}
An equivariant descriptor $\boldsymbol{P}^{\lambda,E}_i$ that automatically satisfies the symmetry of the applied field can then be directly obtained by substituting $V_i$ with $V^{E}_i$ in Eq.~\eqref{eq:lode-descriptor}. Specifically, we find that the field symmetry is manifested in the selection rules $l=\lambda\pm1$ and $m=\mu$~\cite{suppmat}. 
A~ML representation of $n_e$ able to reproduce the non-local charge-density response can be finally constructed via a suitable coupling between $\boldsymbol{P}^{\lambda,E}_i$ and the LODE descriptor of Eq.~\eqref{eq:lode-descriptor}:
\begin{equation}\label{eq:E-descriptor}
    \tilde{\boldsymbol{P}}^{\lambda,E}_i \equiv \left( \boldsymbol{P}^{\lambda}_i\oplus\boldsymbol{P}^{\lambda,E}_i\right)\otimes \boldsymbol{P}^{0}_i\, .
\end{equation}
Note that these algebraic operations account for taking simple sums and products of the individual kernels, thus keeping the dimensionality of the learning problem unchanged with respect to the no-field case (kernel trick~\cite{Konstantinos2009}). A detailed derivation is reported in the SM~\cite{suppmat}. 

\begin{figure}[t!]
    \centering
    \includegraphics[width=8.8cm]{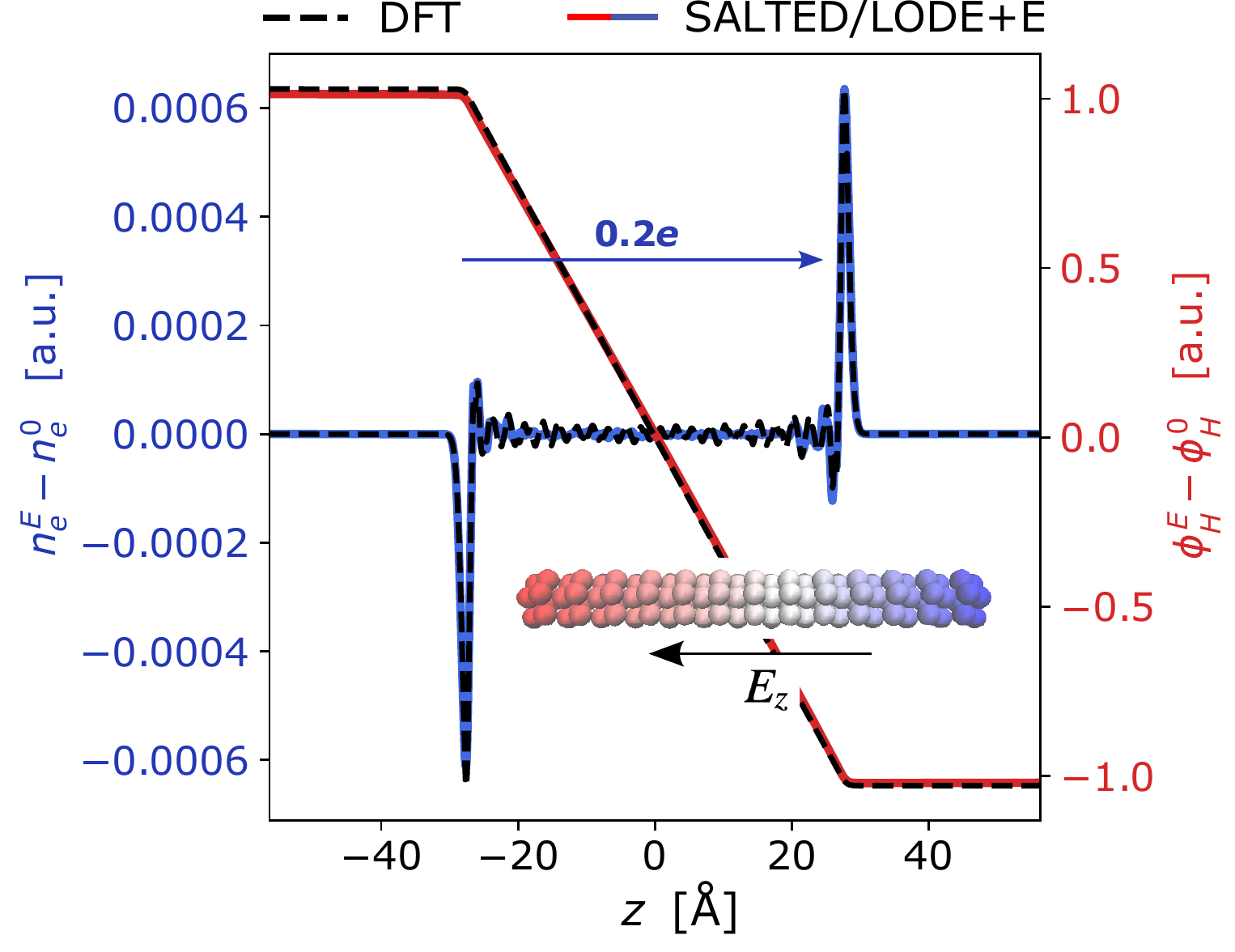}
    \caption{{Extrapolated density and potential response averaged over the $xy$ plane of a Au(100) electrode with 27 metal layers under an applied electric field $E_z=-1$V/{\AA}. Blue line: electron density response. Red line: Hartree potential response. Black dashed line: reference DFT response. Predictions are obtained from a field-dependent SALTED/LODE+E model trained on a dataset which includes electrodes with $n_\text{L}^\text{max}$=15. Inset: representation of the predicted $\phi_\text{H}^E-\phi_\text{H}^0$ through the metal electrode; color code from blue to red corresponds to a linear scale from $-$1.0 to +1.0 a.u., respectively.}} 
    \label{fig:fig2}
\end{figure}

To test the capability of the previously discussed model to predict the electronic charge transfer induced by the applied electric field, we now repeat the size extrapolation exercise already carried out in the previous example, so that the same training and test gold structures are used. 
{Fig.~\ref{fig:fig2} reports the predicted charge-density response for the largest test electrode considered, as compared with the corresponding DFT profile. In spite of the highly extrapolative regime, our finite-field extension of LODE allows us to accurately reproduce the accumulation/depletion of opposite electronic charge at the two sides of the metal electrode, predicting a charge transfer of 0.2$e$ that is in perfect agreement with that of DFT.
To corroborate these results, we report in the same Figure the variation of the Hartree potential, $\Delta \phi_\text{H}$, which can be directly computed from the predicted $\Delta n_e$~\cite{suppmat}.} We find that the expected potential drop of $\sim$2 a.u. between the two metal surfaces is accurately predicted, reproducing the linear decrease of $\Delta \phi_\text{H}$ in the metallic bulk in order to perfectly screen the opposite increase of the external potential $\phi_\text{ext}=-E_z z$. %

\begin{figure}[t!]
    \centering
    \includegraphics[width=8.7cm]{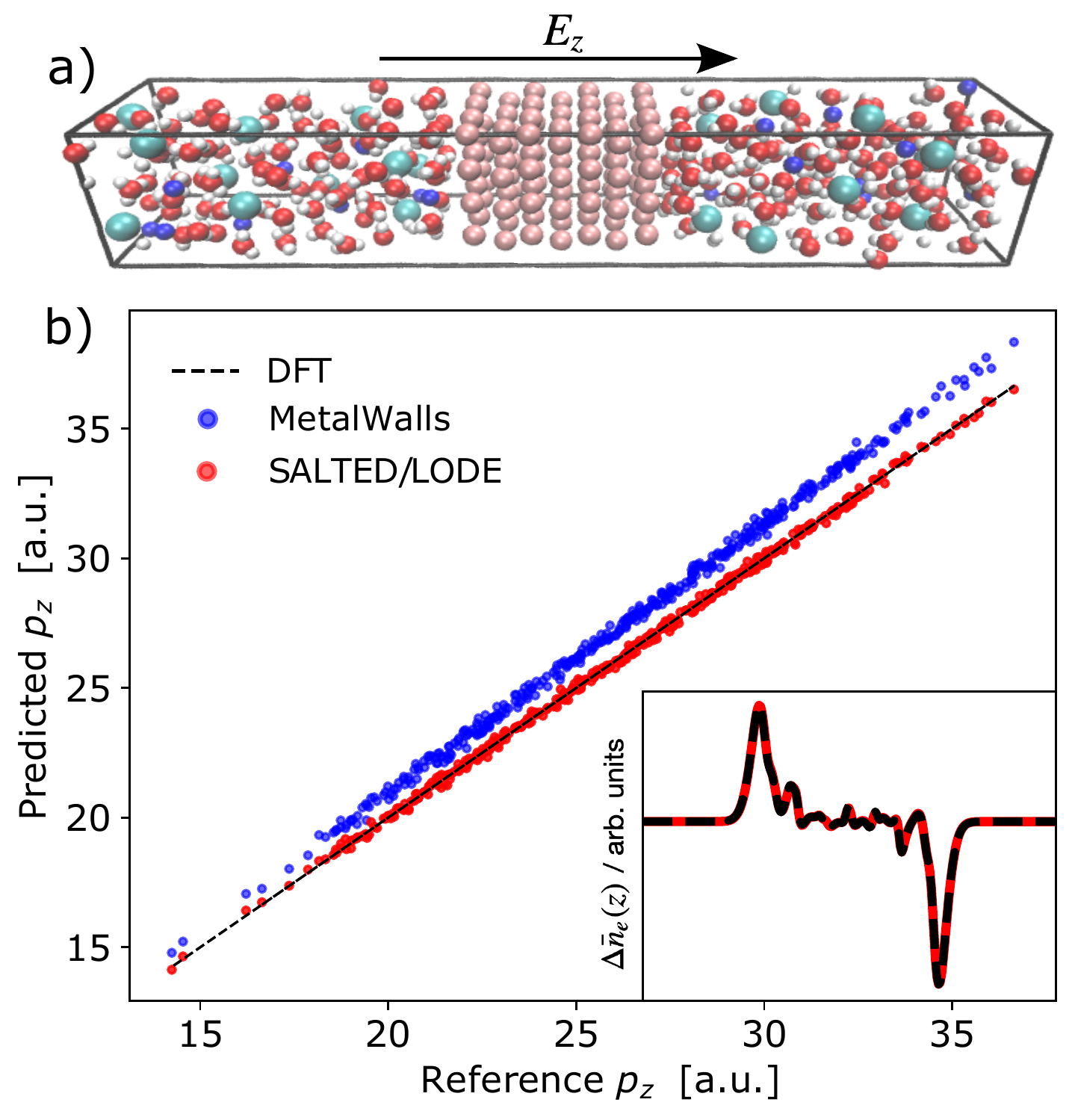}
    \caption{{a) Representation of the physical system under study: a gold electrode is put in contact with a water/NaCl solution under an applied electric field $E_z=0.016$ V/{\AA}, corresponding to a potential drop of $-$1.0 V through the simulation cell~\cite{Dufils2019}.\\ b) Electrode polarization $p_z$ as predicted from the electron-density response induced by 400 electrolyte configurations. Predictions are reported against the reference QM/MM values. Black dashed line: DFT. Blue dots: classical MetalWalls result~\cite{Marin-Lafleche2020}. Red dots: SALTED/LODE results. Inset: predicted electron-density response averaged over the $xy$-plane for a representative electrode/electrolyte configuration (red line), as compared with the reference DFT profile.}} 
    \label{fig:fig3}
\end{figure}

{We continue by showcasing an example where a rigid Au(100) electrode made of 4 unit cell repetitions along the $xy$ plane and 7 metal layers is put in contact with a concentrated water/NaCl solution under a uniform electric field $E_z$, Fig.~\ref{fig:fig3}-a). Following Ref.~\cite{Dufils2019}, a similar setup can be used with 3D periodic boundary conditions to simulate an ionic capacitor under an applied voltage $\Delta V = -E_zL_z$, with $L_z$ the length of the simulation box. In this example, an electric field $E_z=0.016$~V/{\AA} is applied to represent an ionic capacitor subject to a potential difference  of $\Delta V =-1.0$~V. To generate electrolyte configurations that are representative of the given ensemble at $T=298$~K, we run finite-$E$ classical molecular dynamics~\cite{Zhang2020} over 10 ns, using the MetalWalls simulation program~\cite{Marin-Lafleche2020}. From the trajectory so generated, QM/MM calculations of the electrode charge density under the applied electric field $E_z$ are then performed for 2000 uncorrelated configurations. In particular, we treat the gold slab at the same DFT level of theory already adopted in the previous examples, while representing the aqueous electrolyte via classical Gaussian charges~\cite{Laino2006}. Note that because of the classical nature of the electrolyte, setting the electrode in the middle of the simulation box is enough to avoid problems related to the discontinuous jump of the external potential at $z=L_z$. The difference between the QM/MM density and the density of the isolated electrode under $E_z$ is finally considered as a learning target of the SALTED problem.}

{We select 1600 random configurations for training and retain the remaining 400 for testing. SALTED/LODE models are constructed from atom density and potential fields defined from Gaussian widths of $\sigma=0.5${\AA} and $\sigma=1.0${\AA}, respectively, which are both cut off at $r_\text{cut}$=10{\AA}. 
Fig.~\ref{fig:fig3}-b) reports the ML predictions of the derived electrode polarization $p_z$ against the reference DFT values. In so doing, we also compare results obtained from the classical MetalWalls simulation.  We observe that MetalWalls yields a systematic overestimation of the electrode polarization, which accounts for an average deviation of $\langle\Delta p_z\rangle=1.2$~a.u. with respect to DFT. Conversely, our SALTED/LODE model is able to accurately reproduce the charge-density response induced by the field of the various electrolyte configurations.  In particular, the derived ML predictions of $p_z$ present a small root mean square error of 2.9\% of the standard deviation of the polarization vector in the test set.  This is in line with the excellent agreement between the DFT and predicted response averaged over the $xy$-plane, $\Delta \bar{n}(z)$, as reported in the inset of the Figure for a representative test configuration. Performing similar predictions took, on average, $\sim$5 seconds per structure on 128 CPUs versus $\sim$5 minutes of DFT. A substantial speedup of the ML performance will be achieved in the future by relying on a particle mesh Ewald implementation of the LODE potential~\cite{Darden1993}.} 

{We conclude by providing an estimate of the electrical double-layer (EDL) contribution to the differential capacitance of the system, defined from the electrolyte-induced fluctuations of the integrated charge $\pm Q$ accumulated at the two metal surfaces, i.e., $C^\text{EDL}_\text{diff} = \beta \left<\delta Q^2\right>$~\cite{Scalfi2020}. Similarly to the case of $p_z$, the calculation of $Q$ can be performed analytically from the isotropic charge-density components~\cite{suppmat}. Upon predicting $Q$ for 5500 uncorrelated frames of a MetalWalls trajectory of 30ns, we obtain  $C^\text{EDL}_\text{diff}=8.9~\mu$F/cm$^2$. This value is sensibly smaller than what can be obtained from the classical MetalWalls simulation, i.e., $C^\text{EDL}_\text{diff}=10.5 ~\mu$F/cm$^2$, which is found to give larger  fluctuations of the electrode surface charge with respect to the corresponding ML predictions~\cite{suppmat}. By and large, these results highlight the importance of going beyond a classical picture when describing non-local polarization effects in finite conducting materials.}

The presented study shows how the interplay of equivariant, long-range and finite-field learning models can be used to accurately predict the quantum-mechanical response of metal electrodes under electric fields of different nature. 
{In perspective, the ready access to the charge-density response  can provide a rigorous pathway to accurately compute the polarization energy of generic electrochemical interfaces, while having the permanent electrostatics accurately incorporated via local ML models of the electron density~\cite{lewis+21jctc}.}
In this context, an application of major impact will consist in  driving the long-range dynamics of the system at a ML/MM level of theory~\cite{DiPasquale2021}. Finally, we foresee applications of the method in determining the capability of the metal surface to undergo electron-transfer processes~\cite{Lautar2020}, {as well in predicting optical response functions~\cite{Lewis2023} that can serve for a spectroscopic characterization of the interface~\cite{Chowdhury2021}.} 

\begin{acknowledgments}
The authors are grateful to Ari Paavo Seitsonen, Federico Grasselli and Kevin Rossi for useful discussion. AG acknowledges funding from the Swiss National Science Foundation, Grant No. P2ELP2-199747. {This work was supported by the French National Research Agency under the France 2030 program (Grant ANR-22-PEBA-0002). Simulations of the ionic capacitor were performed on the Luxembourg national supercomputer MeluXina (Grant EHPC-REG-2022R02-244).}
\end{acknowledgments}


\begin{thebibliography}{0}%
\makeatletter
\providecommand \@ifxundefined [1]{%
 \@ifx{#1\undefined}
}%
\providecommand \@ifnum [1]{%
 \ifnum #1\expandafter \@firstoftwo
 \else \expandafter \@secondoftwo
 \fi
}%
\providecommand \@ifx [1]{%
 \ifx #1\expandafter \@firstoftwo
 \else \expandafter \@secondoftwo
 \fi
}%
\providecommand \natexlab [1]{#1}%
\providecommand \enquote  [1]{``#1''}%
\providecommand \bibnamefont  [1]{#1}%
\providecommand \bibfnamefont [1]{#1}%
\providecommand \citenamefont [1]{#1}%
\providecommand \href@noop [0]{\@secondoftwo}%
\providecommand \href [0]{\begingroup \@sanitize@url \@href}%
\providecommand \@href[1]{\@@startlink{#1}\@@href}%
\providecommand \@@href[1]{\endgroup#1\@@endlink}%
\providecommand \@sanitize@url [0]{\catcode `\\12\catcode `\$12\catcode
  `\&12\catcode `\#12\catcode `\^12\catcode `\_12\catcode `\%12\relax}%
\providecommand \@@startlink[1]{}%
\providecommand \@@endlink[0]{}%
\providecommand \url  [0]{\begingroup\@sanitize@url \@url }%
\providecommand \@url [1]{\endgroup\@href {#1}{\urlprefix }}%
\providecommand \urlprefix  [0]{URL }%
\providecommand \Eprint [0]{\href }%
\providecommand \doibase [0]{http://dx.doi.org/}%
\providecommand \selectlanguage [0]{\@gobble}%
\providecommand \bibinfo  [0]{\@secondoftwo}%
\providecommand \bibfield  [0]{\@secondoftwo}%
\providecommand \translation [1]{[#1]}%
\providecommand \BibitemOpen [0]{}%
\providecommand \bibitemStop [0]{}%
\providecommand \bibitemNoStop [0]{.\EOS\space}%
\providecommand \EOS [0]{\spacefactor3000\relax}%
\providecommand \BibitemShut  [1]{\csname bibitem#1\endcsname}%
\let\auto@bib@innerbib\@empty
\end{thebibliography}%


\begin{thebibliography}{73}%
\makeatletter
\providecommand \@ifxundefined [1]{%
 \@ifx{#1\undefined}
}%
\providecommand \@ifnum [1]{%
 \ifnum #1\expandafter \@firstoftwo
 \else \expandafter \@secondoftwo
 \fi
}%
\providecommand \@ifx [1]{%
 \ifx #1\expandafter \@firstoftwo
 \else \expandafter \@secondoftwo
 \fi
}%
\providecommand \natexlab [1]{#1}%
\providecommand \enquote  [1]{``#1''}%
\providecommand \bibnamefont  [1]{#1}%
\providecommand \bibfnamefont [1]{#1}%
\providecommand \citenamefont [1]{#1}%
\providecommand \href@noop [0]{\@secondoftwo}%
\providecommand \href [0]{\begingroup \@sanitize@url \@href}%
\providecommand \@href[1]{\@@startlink{#1}\@@href}%
\providecommand \@@href[1]{\endgroup#1\@@endlink}%
\providecommand \@sanitize@url [0]{\catcode `\\12\catcode `\$12\catcode
  `\&12\catcode `\#12\catcode `\^12\catcode `\_12\catcode `\%12\relax}%
\providecommand \@@startlink[1]{}%
\providecommand \@@endlink[0]{}%
\providecommand \url  [0]{\begingroup\@sanitize@url \@url }%
\providecommand \@url [1]{\endgroup\@href {#1}{\urlprefix }}%
\providecommand \urlprefix  [0]{URL }%
\providecommand \Eprint [0]{\href }%
\providecommand \doibase [0]{http://dx.doi.org/}%
\providecommand \selectlanguage [0]{\@gobble}%
\providecommand \bibinfo  [0]{\@secondoftwo}%
\providecommand \bibfield  [0]{\@secondoftwo}%
\providecommand \translation [1]{[#1]}%
\providecommand \BibitemOpen [0]{}%
\providecommand \bibitemStop [0]{}%
\providecommand \bibitemNoStop [0]{.\EOS\space}%
\providecommand \EOS [0]{\spacefactor3000\relax}%
\providecommand \BibitemShut  [1]{\csname bibitem#1\endcsname}%
\let\auto@bib@innerbib\@empty
\bibitem [{\citenamefont {Lautar}\ \emph
  {et~al.}(2020{\natexlab{a}})\citenamefont {Lautar}, \citenamefont {Bitenc},
  \citenamefont {Rejec}, \citenamefont {Dominko}, \citenamefont {Filhol},\ and\
  \citenamefont {Doublet}}]{Lautar2020jacs}%
  \BibitemOpen
  \bibfield  {author} {\bibinfo {author} {\bibfnamefont {A.~K.}\ \bibnamefont
  {Lautar}}, \bibinfo {author} {\bibfnamefont {J.}~\bibnamefont {Bitenc}},
  \bibinfo {author} {\bibfnamefont {T.}~\bibnamefont {Rejec}}, \bibinfo
  {author} {\bibfnamefont {R.}~\bibnamefont {Dominko}}, \bibinfo {author}
  {\bibfnamefont {J.-S.}\ \bibnamefont {Filhol}}, \ and\ \bibinfo {author}
  {\bibfnamefont {M.-L.}\ \bibnamefont {Doublet}},\ }\href {\doibase
  10.1021/jacs.9b12474} {\bibfield  {journal} {\bibinfo  {journal} {Journal of
  the American Chemical Society}\ }\textbf {\bibinfo {volume} {142}},\ \bibinfo
  {pages} {5146} (\bibinfo {year} {2020}{\natexlab{a}})}\BibitemShut {NoStop}%
\bibitem [{\citenamefont {Beinlich}\ \emph {et~al.}(2022)\citenamefont
  {Beinlich}, \citenamefont {H\"ormann},\ and\ \citenamefont
  {Reuter}}]{Beinlich2022}%
  \BibitemOpen
  \bibfield  {author} {\bibinfo {author} {\bibfnamefont {S.~D.}\ \bibnamefont
  {Beinlich}}, \bibinfo {author} {\bibfnamefont {N.~G.}\ \bibnamefont
  {H\"ormann}}, \ and\ \bibinfo {author} {\bibfnamefont {K.}~\bibnamefont
  {Reuter}},\ }\href {\doibase 10.1021/acscatal.2c00997} {\bibfield  {journal}
  {\bibinfo  {journal} {ACS Catalysis}\ }\textbf {\bibinfo {volume} {12}},\
  \bibinfo {pages} {6143} (\bibinfo {year} {2022})}\BibitemShut {NoStop}%
\bibitem [{\citenamefont {Karmodak}\ \emph {et~al.}(2022)\citenamefont
  {Karmodak}, \citenamefont {Bursi},\ and\ \citenamefont
  {Andreussi}}]{Karmodak2022}%
  \BibitemOpen
  \bibfield  {author} {\bibinfo {author} {\bibfnamefont {N.}~\bibnamefont
  {Karmodak}}, \bibinfo {author} {\bibfnamefont {L.}~\bibnamefont {Bursi}}, \
  and\ \bibinfo {author} {\bibfnamefont {O.}~\bibnamefont {Andreussi}},\ }\href
  {\doibase 10.1021/acs.jpclett.1c03431} {\bibfield  {journal} {\bibinfo
  {journal} {The Journal of Physical Chemistry Letters}\ }\textbf {\bibinfo
  {volume} {13}},\ \bibinfo {pages} {58} (\bibinfo {year} {2022})}\BibitemShut
  {NoStop}%
\bibitem [{\citenamefont {N\'orskov}\ \emph {et~al.}(2011)\citenamefont
  {N\'orskov}, \citenamefont {Abild-Pedersen}, \citenamefont {Studt},\ and\
  \citenamefont {Bligaard}}]{norskov2011}%
  \BibitemOpen
  \bibfield  {author} {\bibinfo {author} {\bibfnamefont {J.~K.}\ \bibnamefont
  {N\'orskov}}, \bibinfo {author} {\bibfnamefont {F.}~\bibnamefont
  {Abild-Pedersen}}, \bibinfo {author} {\bibfnamefont {F.}~\bibnamefont
  {Studt}}, \ and\ \bibinfo {author} {\bibfnamefont {T.}~\bibnamefont
  {Bligaard}},\ }\href {\doibase 10.1073/pnas.1006652108} {\bibfield  {journal}
  {\bibinfo  {journal} {Proceedings of the National Academy of Sciences}\
  }\textbf {\bibinfo {volume} {108}},\ \bibinfo {pages} {937} (\bibinfo {year}
  {2011})}\BibitemShut {NoStop}%
\bibitem [{\citenamefont {Grosjean}\ \emph {et~al.}(2019)\citenamefont
  {Grosjean}, \citenamefont {Bocquet},\ and\ \citenamefont
  {Vuilleumier}}]{Grosjean2019}%
  \BibitemOpen
  \bibfield  {author} {\bibinfo {author} {\bibfnamefont {B.}~\bibnamefont
  {Grosjean}}, \bibinfo {author} {\bibfnamefont {M.-L.}\ \bibnamefont
  {Bocquet}}, \ and\ \bibinfo {author} {\bibfnamefont {R.}~\bibnamefont
  {Vuilleumier}},\ }\href {https://doi.org/10.1038/s41467-019-09708-7}
  {\bibfield  {journal} {\bibinfo  {journal} {Nature Communications}\ }\textbf
  {\bibinfo {volume} {10}},\ \bibinfo {pages} {1656} (\bibinfo {year}
  {2019})}\BibitemShut {NoStop}%
\bibitem [{\citenamefont {Gerrits}\ \emph {et~al.}(2020)\citenamefont
  {Gerrits}, \citenamefont {Smeets}, \citenamefont {Vuckovic}, \citenamefont
  {Powell}, \citenamefont {Doblhoff-Dier},\ and\ \citenamefont
  {Kroes}}]{Gerrits2020}%
  \BibitemOpen
  \bibfield  {author} {\bibinfo {author} {\bibfnamefont {N.}~\bibnamefont
  {Gerrits}}, \bibinfo {author} {\bibfnamefont {E.~W.~F.}\ \bibnamefont
  {Smeets}}, \bibinfo {author} {\bibfnamefont {S.}~\bibnamefont {Vuckovic}},
  \bibinfo {author} {\bibfnamefont {A.~D.}\ \bibnamefont {Powell}}, \bibinfo
  {author} {\bibfnamefont {K.}~\bibnamefont {Doblhoff-Dier}}, \ and\ \bibinfo
  {author} {\bibfnamefont {G.-J.}\ \bibnamefont {Kroes}},\ }\href {\doibase
  10.1021/acs.jpclett.0c02452} {\bibfield  {journal} {\bibinfo  {journal} {The
  Journal of Physical Chemistry Letters}\ }\textbf {\bibinfo {volume} {11}},\
  \bibinfo {pages} {10552} (\bibinfo {year} {2020})}\BibitemShut {NoStop}%
\bibitem [{\citenamefont {Souza}\ \emph {et~al.}(2013)\citenamefont {Souza},
  \citenamefont {Rungger}, \citenamefont {Pemmaraju}, \citenamefont
  {Schwingenschloegl},\ and\ \citenamefont {Sanvito}}]{Souza2013}%
  \BibitemOpen
  \bibfield  {author} {\bibinfo {author} {\bibfnamefont {A.~M.}\ \bibnamefont
  {Souza}}, \bibinfo {author} {\bibfnamefont {I.}~\bibnamefont {Rungger}},
  \bibinfo {author} {\bibfnamefont {C.~D.}\ \bibnamefont {Pemmaraju}}, \bibinfo
  {author} {\bibfnamefont {U.}~\bibnamefont {Schwingenschloegl}}, \ and\
  \bibinfo {author} {\bibfnamefont {S.}~\bibnamefont {Sanvito}},\ }\href
  {\doibase 10.1103/PhysRevB.88.165112} {\bibfield  {journal} {\bibinfo
  {journal} {Phys. Rev. B}\ }\textbf {\bibinfo {volume} {88}},\ \bibinfo
  {pages} {165112} (\bibinfo {year} {2013})}\BibitemShut {NoStop}%
\bibitem [{\citenamefont {Freysoldt}\ \emph {et~al.}(2020)\citenamefont
  {Freysoldt}, \citenamefont {Mishra}, \citenamefont {Ashton},\ and\
  \citenamefont {Neugebauer}}]{Freysoldt2020}%
  \BibitemOpen
  \bibfield  {author} {\bibinfo {author} {\bibfnamefont {C.}~\bibnamefont
  {Freysoldt}}, \bibinfo {author} {\bibfnamefont {A.}~\bibnamefont {Mishra}},
  \bibinfo {author} {\bibfnamefont {M.}~\bibnamefont {Ashton}}, \ and\ \bibinfo
  {author} {\bibfnamefont {J.}~\bibnamefont {Neugebauer}},\ }\href {\doibase
  10.1103/PhysRevB.102.045403} {\bibfield  {journal} {\bibinfo  {journal}
  {Phys. Rev. B}\ }\textbf {\bibinfo {volume} {102}},\ \bibinfo {pages}
  {045403} (\bibinfo {year} {2020})}\BibitemShut {NoStop}%
\bibitem [{\citenamefont {Lozovoi}\ \emph {et~al.}(2001)\citenamefont
  {Lozovoi}, \citenamefont {Alavi}, \citenamefont {Kohanoff},\ and\
  \citenamefont {Lynden-Bell}}]{Lozovoi2001}%
  \BibitemOpen
  \bibfield  {author} {\bibinfo {author} {\bibfnamefont {A.~Y.}\ \bibnamefont
  {Lozovoi}}, \bibinfo {author} {\bibfnamefont {A.}~\bibnamefont {Alavi}},
  \bibinfo {author} {\bibfnamefont {J.}~\bibnamefont {Kohanoff}}, \ and\
  \bibinfo {author} {\bibfnamefont {R.~M.}\ \bibnamefont {Lynden-Bell}},\
  }\href {\doibase 10.1063/1.1379327} {\bibfield  {journal} {\bibinfo
  {journal} {The Journal of Chemical Physics}\ }\textbf {\bibinfo {volume}
  {115}},\ \bibinfo {pages} {1661} (\bibinfo {year} {2001})}\BibitemShut
  {NoStop}%
\bibitem [{\citenamefont {Filhol}\ and\ \citenamefont
  {Neurock}(2006)}]{Filhol2006}%
  \BibitemOpen
  \bibfield  {author} {\bibinfo {author} {\bibfnamefont {J.-S.}\ \bibnamefont
  {Filhol}}\ and\ \bibinfo {author} {\bibfnamefont {M.}~\bibnamefont
  {Neurock}},\ }\href {https://doi.org/10.1002/anie.200502540} {\bibfield
  {journal} {\bibinfo  {journal} {Angewandte Chemie International Edition}\
  }\textbf {\bibinfo {volume} {45}},\ \bibinfo {pages} {402} (\bibinfo {year}
  {2006})}\BibitemShut {NoStop}%
\bibitem [{\citenamefont {Letchworth-Weaver}\ and\ \citenamefont
  {Arias}(2012)}]{Letchworth2012}%
  \BibitemOpen
  \bibfield  {author} {\bibinfo {author} {\bibfnamefont {K.}~\bibnamefont
  {Letchworth-Weaver}}\ and\ \bibinfo {author} {\bibfnamefont {T.~A.}\
  \bibnamefont {Arias}},\ }\href {\doibase 10.1103/PhysRevB.86.075140}
  {\bibfield  {journal} {\bibinfo  {journal} {Phys. Rev. B}\ }\textbf {\bibinfo
  {volume} {86}},\ \bibinfo {pages} {075140} (\bibinfo {year}
  {2012})}\BibitemShut {NoStop}%
\bibitem [{\citenamefont {H\"ormann}\ \emph {et~al.}(2019)\citenamefont
  {H\"ormann}, \citenamefont {Andreussi},\ and\ \citenamefont
  {Marzari}}]{Hoermann2019}%
  \BibitemOpen
  \bibfield  {author} {\bibinfo {author} {\bibfnamefont {N.~G.}\ \bibnamefont
  {H\"ormann}}, \bibinfo {author} {\bibfnamefont {O.}~\bibnamefont
  {Andreussi}}, \ and\ \bibinfo {author} {\bibfnamefont {N.}~\bibnamefont
  {Marzari}},\ }\href {https://doi.org/10.1063/1.5054580} {\bibfield  {journal}
  {\bibinfo  {journal} {The Journal of Chemical Physics}\ }\textbf {\bibinfo
  {volume} {150}},\ \bibinfo {pages} {041730} (\bibinfo {year}
  {2019})}\BibitemShut {NoStop}%
\bibitem [{\citenamefont {Melander}\ \emph {et~al.}(2019)\citenamefont
  {Melander}, \citenamefont {Kuisma}, \citenamefont {Christensen},\ and\
  \citenamefont {Honkala}}]{Melander2019}%
  \BibitemOpen
  \bibfield  {author} {\bibinfo {author} {\bibfnamefont {M.~M.}\ \bibnamefont
  {Melander}}, \bibinfo {author} {\bibfnamefont {M.~J.}\ \bibnamefont
  {Kuisma}}, \bibinfo {author} {\bibfnamefont {T.~E.~K.}\ \bibnamefont
  {Christensen}}, \ and\ \bibinfo {author} {\bibfnamefont {K.}~\bibnamefont
  {Honkala}},\ }\href {https://doi.org/10.1063/1.5047829} {\bibfield  {journal}
  {\bibinfo  {journal} {The Journal of Chemical Physics}\ }\textbf {\bibinfo
  {volume} {150}},\ \bibinfo {pages} {041706} (\bibinfo {year}
  {2019})}\BibitemShut {NoStop}%
\bibitem [{\citenamefont {Domínguez-Flores}\ and\ \citenamefont
  {Melander}(2023)}]{Dominguez2023}%
  \BibitemOpen
  \bibfield  {author} {\bibinfo {author} {\bibfnamefont {F.}~\bibnamefont
  {Domínguez-Flores}}\ and\ \bibinfo {author} {\bibfnamefont {M.~M.}\
  \bibnamefont {Melander}},\ }\href {\doibase 10.1063/5.0138197} {\bibfield
  {journal} {\bibinfo  {journal} {The Journal of Chemical Physics}\ }\textbf
  {\bibinfo {volume} {158}},\ \bibinfo {pages} {144701} (\bibinfo {year}
  {2023})}\BibitemShut {NoStop}%
\bibitem [{\citenamefont {Stengel}\ and\ \citenamefont
  {Spaldin}(2007)}]{Stengel2007}%
  \BibitemOpen
  \bibfield  {author} {\bibinfo {author} {\bibfnamefont {M.}~\bibnamefont
  {Stengel}}\ and\ \bibinfo {author} {\bibfnamefont {N.~A.}\ \bibnamefont
  {Spaldin}},\ }\href {\doibase 10.1103/PhysRevB.75.205121} {\bibfield
  {journal} {\bibinfo  {journal} {Phys. Rev. B}\ }\textbf {\bibinfo {volume}
  {75}},\ \bibinfo {pages} {205121} (\bibinfo {year} {2007})}\BibitemShut
  {NoStop}%
\bibitem [{\citenamefont {Che}\ \emph {et~al.}(2017)\citenamefont {Che},
  \citenamefont {Gray}, \citenamefont {Ha},\ and\ \citenamefont
  {McEwen}}]{Che2017}%
  \BibitemOpen
  \bibfield  {author} {\bibinfo {author} {\bibfnamefont {F.}~\bibnamefont
  {Che}}, \bibinfo {author} {\bibfnamefont {J.~T.}\ \bibnamefont {Gray}},
  \bibinfo {author} {\bibfnamefont {S.}~\bibnamefont {Ha}}, \ and\ \bibinfo
  {author} {\bibfnamefont {J.-S.}\ \bibnamefont {McEwen}},\ }\href {\doibase
  10.1021/acscatal.6b02318} {\bibfield  {journal} {\bibinfo  {journal} {ACS
  Catalysis}\ }\textbf {\bibinfo {volume} {7}},\ \bibinfo {pages} {551}
  (\bibinfo {year} {2017})}\BibitemShut {NoStop}%
\bibitem [{\citenamefont {Ke}\ \emph {et~al.}(2022)\citenamefont {Ke},
  \citenamefont {Lin},\ and\ \citenamefont {Liu}}]{Ke2022}%
  \BibitemOpen
  \bibfield  {author} {\bibinfo {author} {\bibfnamefont {C.}~\bibnamefont
  {Ke}}, \bibinfo {author} {\bibfnamefont {Z.}~\bibnamefont {Lin}}, \ and\
  \bibinfo {author} {\bibfnamefont {S.}~\bibnamefont {Liu}},\ }\href {\doibase
  10.1021/acscatal.2c04961} {\bibfield  {journal} {\bibinfo  {journal} {ACS
  Catalysis}\ }\textbf {\bibinfo {volume} {12}},\ \bibinfo {pages} {13542}
  (\bibinfo {year} {2022})}\BibitemShut {NoStop}%
\bibitem [{\citenamefont {Khatib}\ \emph {et~al.}(2021)\citenamefont {Khatib},
  \citenamefont {Kumar}, \citenamefont {Sanvito}, \citenamefont {Sulpizi},\
  and\ \citenamefont {Cucinotta}}]{Khatib2021}%
  \BibitemOpen
  \bibfield  {author} {\bibinfo {author} {\bibfnamefont {R.}~\bibnamefont
  {Khatib}}, \bibinfo {author} {\bibfnamefont {A.}~\bibnamefont {Kumar}},
  \bibinfo {author} {\bibfnamefont {S.}~\bibnamefont {Sanvito}}, \bibinfo
  {author} {\bibfnamefont {M.}~\bibnamefont {Sulpizi}}, \ and\ \bibinfo
  {author} {\bibfnamefont {C.~S.}\ \bibnamefont {Cucinotta}},\ }\href {\doibase
  https://doi.org/10.1016/j.electacta.2021.138875} {\bibfield  {journal}
  {\bibinfo  {journal} {Electrochimica Acta}\ }\textbf {\bibinfo {volume}
  {391}},\ \bibinfo {pages} {138875} (\bibinfo {year} {2021})}\BibitemShut
  {NoStop}%
\bibitem [{\citenamefont {Chen}\ \emph {et~al.}(2022)\citenamefont {Chen},
  \citenamefont {Le}, \citenamefont {Kuang},\ and\ \citenamefont
  {Cheng}}]{Chen2022}%
  \BibitemOpen
  \bibfield  {author} {\bibinfo {author} {\bibfnamefont {A.}~\bibnamefont
  {Chen}}, \bibinfo {author} {\bibfnamefont {J.-B.}\ \bibnamefont {Le}},
  \bibinfo {author} {\bibfnamefont {Y.}~\bibnamefont {Kuang}}, \ and\ \bibinfo
  {author} {\bibfnamefont {J.}~\bibnamefont {Cheng}},\ }\href {\doibase
  10.1063/5.0100678} {\bibfield  {journal} {\bibinfo  {journal} {The Journal of
  Chemical Physics}\ }\textbf {\bibinfo {volume} {157}},\ \bibinfo {pages}
  {094702} (\bibinfo {year} {2022})}\BibitemShut {NoStop}%
\bibitem [{\citenamefont {Jeanmairet}\ \emph {et~al.}(2022)\citenamefont
  {Jeanmairet}, \citenamefont {Rotenberg},\ and\ \citenamefont
  {Salanne}}]{Jeanmairet2022}%
  \BibitemOpen
  \bibfield  {author} {\bibinfo {author} {\bibfnamefont {G.}~\bibnamefont
  {Jeanmairet}}, \bibinfo {author} {\bibfnamefont {B.}~\bibnamefont
  {Rotenberg}}, \ and\ \bibinfo {author} {\bibfnamefont {M.}~\bibnamefont
  {Salanne}},\ }\href {\doibase 10.1021/acs.chemrev.1c00925} {\bibfield
  {journal} {\bibinfo  {journal} {Chemical Reviews}\ }\textbf {\bibinfo
  {volume} {122}},\ \bibinfo {pages} {10860} (\bibinfo {year}
  {2022})}\BibitemShut {NoStop}%
\bibitem [{\citenamefont {Siepmann}\ and\ \citenamefont
  {Sprik}(1995)}]{Siepmann1995}%
  \BibitemOpen
  \bibfield  {author} {\bibinfo {author} {\bibfnamefont {J.~I.}\ \bibnamefont
  {Siepmann}}\ and\ \bibinfo {author} {\bibfnamefont {M.}~\bibnamefont
  {Sprik}},\ }\href {\doibase 10.1063/1.469429} {\bibfield  {journal} {\bibinfo
   {journal} {The Journal of Chemical Physics}\ }\textbf {\bibinfo {volume}
  {102}},\ \bibinfo {pages} {511} (\bibinfo {year} {1995})}\BibitemShut
  {NoStop}%
\bibitem [{\citenamefont {Reed}\ \emph {et~al.}(2007)\citenamefont {Reed},
  \citenamefont {Lanning},\ and\ \citenamefont {Madden}}]{Reed2007}%
  \BibitemOpen
  \bibfield  {author} {\bibinfo {author} {\bibfnamefont {S.~K.}\ \bibnamefont
  {Reed}}, \bibinfo {author} {\bibfnamefont {O.~J.}\ \bibnamefont {Lanning}}, \
  and\ \bibinfo {author} {\bibfnamefont {P.~A.}\ \bibnamefont {Madden}},\
  }\href {https://doi.org/10.1063/1.2464084} {\bibfield  {journal} {\bibinfo
  {journal} {The Journal of Chemical Physics}\ }\textbf {\bibinfo {volume}
  {126}},\ \bibinfo {pages} {084704} (\bibinfo {year} {2007})}\BibitemShut
  {NoStop}%
\bibitem [{\citenamefont {Reed}\ \emph {et~al.}(2008)\citenamefont {Reed},
  \citenamefont {Madden},\ and\ \citenamefont {Papadopoulos}}]{Reed2008}%
  \BibitemOpen
  \bibfield  {author} {\bibinfo {author} {\bibfnamefont {S.~K.}\ \bibnamefont
  {Reed}}, \bibinfo {author} {\bibfnamefont {P.~A.}\ \bibnamefont {Madden}}, \
  and\ \bibinfo {author} {\bibfnamefont {A.}~\bibnamefont {Papadopoulos}},\
  }\href {https://doi.org/10.1063/1.2844801} {\bibfield  {journal} {\bibinfo
  {journal} {The Journal of Chemical Physics}\ }\textbf {\bibinfo {volume}
  {128}},\ \bibinfo {pages} {124701} (\bibinfo {year} {2008})}\BibitemShut
  {NoStop}%
\bibitem [{\citenamefont {Limmer}\ \emph {et~al.}(2013)\citenamefont {Limmer},
  \citenamefont {Merlet}, \citenamefont {Salanne}, \citenamefont {Chandler},
  \citenamefont {Madden}, \citenamefont {van Roij},\ and\ \citenamefont
  {Rotenberg}}]{Limmer2013}%
  \BibitemOpen
  \bibfield  {author} {\bibinfo {author} {\bibfnamefont {D.~T.}\ \bibnamefont
  {Limmer}}, \bibinfo {author} {\bibfnamefont {C.}~\bibnamefont {Merlet}},
  \bibinfo {author} {\bibfnamefont {M.}~\bibnamefont {Salanne}}, \bibinfo
  {author} {\bibfnamefont {D.}~\bibnamefont {Chandler}}, \bibinfo {author}
  {\bibfnamefont {P.~A.}\ \bibnamefont {Madden}}, \bibinfo {author}
  {\bibfnamefont {R.}~\bibnamefont {van Roij}}, \ and\ \bibinfo {author}
  {\bibfnamefont {B.}~\bibnamefont {Rotenberg}},\ }\href
  {https://link.aps.org/doi/10.1103/PhysRevLett.111.106102} {\bibfield
  {journal} {\bibinfo  {journal} {Phys. Rev. Lett.}\ }\textbf {\bibinfo
  {volume} {111}},\ \bibinfo {pages} {106102} (\bibinfo {year}
  {2013})}\BibitemShut {NoStop}%
\bibitem [{\citenamefont {Scalfi}\ \emph
  {et~al.}(2020{\natexlab{a}})\citenamefont {Scalfi}, \citenamefont {Limmer},
  \citenamefont {Coretti}, \citenamefont {Bonella}, \citenamefont {Madden},
  \citenamefont {Salanne},\ and\ \citenamefont {Rotenberg}}]{Scalfi2020}%
  \BibitemOpen
  \bibfield  {author} {\bibinfo {author} {\bibfnamefont {L.}~\bibnamefont
  {Scalfi}}, \bibinfo {author} {\bibfnamefont {D.~T.}\ \bibnamefont {Limmer}},
  \bibinfo {author} {\bibfnamefont {A.}~\bibnamefont {Coretti}}, \bibinfo
  {author} {\bibfnamefont {S.}~\bibnamefont {Bonella}}, \bibinfo {author}
  {\bibfnamefont {P.~A.}\ \bibnamefont {Madden}}, \bibinfo {author}
  {\bibfnamefont {M.}~\bibnamefont {Salanne}}, \ and\ \bibinfo {author}
  {\bibfnamefont {B.}~\bibnamefont {Rotenberg}},\ }\href
  {https://doi.org/10.1039/C9CP06285H} {\bibfield  {journal} {\bibinfo
  {journal} {Phys. Chem. Chem. Phys.}\ }\textbf {\bibinfo {volume} {22}},\
  \bibinfo {pages} {10480} (\bibinfo {year} {2020}{\natexlab{a}})}\BibitemShut
  {NoStop}%
\bibitem [{\citenamefont {Petersen}\ \emph {et~al.}(2012)\citenamefont
  {Petersen}, \citenamefont {Kumar}, \citenamefont {White},\ and\ \citenamefont
  {Voth}}]{Petersen2012}%
  \BibitemOpen
  \bibfield  {author} {\bibinfo {author} {\bibfnamefont {M.~K.}\ \bibnamefont
  {Petersen}}, \bibinfo {author} {\bibfnamefont {R.}~\bibnamefont {Kumar}},
  \bibinfo {author} {\bibfnamefont {H.~S.}\ \bibnamefont {White}}, \ and\
  \bibinfo {author} {\bibfnamefont {G.~A.}\ \bibnamefont {Voth}},\ }\href
  {https://doi.org/10.1021/jp210252g} {\bibfield  {journal} {\bibinfo
  {journal} {The Journal of Physical Chemistry C}\ }\textbf {\bibinfo {volume}
  {116}},\ \bibinfo {pages} {4903} (\bibinfo {year} {2012})}\BibitemShut
  {NoStop}%
\bibitem [{\citenamefont {Dwelle}\ and\ \citenamefont
  {Willard}(2019)}]{Dwelle2019}%
  \BibitemOpen
  \bibfield  {author} {\bibinfo {author} {\bibfnamefont {K.~A.}\ \bibnamefont
  {Dwelle}}\ and\ \bibinfo {author} {\bibfnamefont {A.~P.}\ \bibnamefont
  {Willard}},\ }\href {https://doi.org/10.1021/acs.jpcc.9b06635} {\bibfield
  {journal} {\bibinfo  {journal} {The Journal of Physical Chemistry C}\
  }\textbf {\bibinfo {volume} {123}},\ \bibinfo {pages} {24095} (\bibinfo
  {year} {2019})}\BibitemShut {NoStop}%
\bibitem [{\citenamefont {Scalfi}\ \emph
  {et~al.}(2020{\natexlab{b}})\citenamefont {Scalfi}, \citenamefont {Dufils},
  \citenamefont {Reeves}, \citenamefont {Rotenberg},\ and\ \citenamefont
  {Salanne}}]{Scalfi2020thomas-fermi}%
  \BibitemOpen
  \bibfield  {author} {\bibinfo {author} {\bibfnamefont {L.}~\bibnamefont
  {Scalfi}}, \bibinfo {author} {\bibfnamefont {T.}~\bibnamefont {Dufils}},
  \bibinfo {author} {\bibfnamefont {K.~G.}\ \bibnamefont {Reeves}}, \bibinfo
  {author} {\bibfnamefont {B.}~\bibnamefont {Rotenberg}}, \ and\ \bibinfo
  {author} {\bibfnamefont {M.}~\bibnamefont {Salanne}},\ }\href {\doibase
  10.1063/5.0028232} {\bibfield  {journal} {\bibinfo  {journal} {The Journal of
  Chemical Physics}\ }\textbf {\bibinfo {volume} {153}},\ \bibinfo {pages}
  {174704} (\bibinfo {year} {2020}{\natexlab{b}})}\BibitemShut {NoStop}%
\bibitem [{\citenamefont {Scalfi}\ and\ \citenamefont
  {Rotenberg}(2021)}]{Scalfi2021}%
  \BibitemOpen
  \bibfield  {author} {\bibinfo {author} {\bibfnamefont {L.}~\bibnamefont
  {Scalfi}}\ and\ \bibinfo {author} {\bibfnamefont {B.}~\bibnamefont
  {Rotenberg}},\ }\href {\doibase 10.1073/pnas.2108769118} {\bibfield
  {journal} {\bibinfo  {journal} {Proceedings of the National Academy of
  Sciences}\ }\textbf {\bibinfo {volume} {118}},\ \bibinfo {pages}
  {e2108769118} (\bibinfo {year} {2021})}\BibitemShut {NoStop}%
\bibitem [{\citenamefont {Di~Pasquale}\ \emph {et~al.}(2021)\citenamefont
  {Di~Pasquale}, \citenamefont {Elliott}, \citenamefont {Hadjidoukas},\ and\
  \citenamefont {Carbone}}]{DiPasquale2021}%
  \BibitemOpen
  \bibfield  {author} {\bibinfo {author} {\bibfnamefont {N.}~\bibnamefont
  {Di~Pasquale}}, \bibinfo {author} {\bibfnamefont {J.~D.}\ \bibnamefont
  {Elliott}}, \bibinfo {author} {\bibfnamefont {P.}~\bibnamefont
  {Hadjidoukas}}, \ and\ \bibinfo {author} {\bibfnamefont {P.}~\bibnamefont
  {Carbone}},\ }\href {\doibase 10.1021/acs.jctc.1c00360} {\bibfield  {journal}
  {\bibinfo  {journal} {Journal of Chemical Theory and Computation}\ }\textbf
  {\bibinfo {volume} {17}},\ \bibinfo {pages} {4477} (\bibinfo {year}
  {2021})}\BibitemShut {NoStop}%
\bibitem [{\citenamefont {Di~Pasquale}\ \emph {et~al.}(2023)\citenamefont
  {Di~Pasquale}, \citenamefont {Finney}, \citenamefont {Elliott}, \citenamefont
  {Carbone},\ and\ \citenamefont {Salvalaglio}}]{DiPasquale2023}%
  \BibitemOpen
  \bibfield  {author} {\bibinfo {author} {\bibfnamefont {N.}~\bibnamefont
  {Di~Pasquale}}, \bibinfo {author} {\bibfnamefont {A.~R.}\ \bibnamefont
  {Finney}}, \bibinfo {author} {\bibfnamefont {J.~D.}\ \bibnamefont {Elliott}},
  \bibinfo {author} {\bibfnamefont {P.}~\bibnamefont {Carbone}}, \ and\
  \bibinfo {author} {\bibfnamefont {M.}~\bibnamefont {Salvalaglio}},\ }\href
  {\doibase 10.1063/5.0138267} {\bibfield  {journal} {\bibinfo  {journal} {The
  Journal of Chemical Physics}\ }\textbf {\bibinfo {volume} {158}},\ \bibinfo
  {pages} {134714} (\bibinfo {year} {2023})}\BibitemShut {NoStop}%
\bibitem [{\citenamefont {Brockherde}\ \emph {et~al.}(2017)\citenamefont
  {Brockherde}, \citenamefont {Vogt}, \citenamefont {Li}, \citenamefont
  {Tuckerman}, \citenamefont {Burke},\ and\ \citenamefont
  {M{\"u}ller}}]{broc+17nc}%
  \BibitemOpen
  \bibfield  {author} {\bibinfo {author} {\bibfnamefont {F.}~\bibnamefont
  {Brockherde}}, \bibinfo {author} {\bibfnamefont {L.}~\bibnamefont {Vogt}},
  \bibinfo {author} {\bibfnamefont {L.}~\bibnamefont {Li}}, \bibinfo {author}
  {\bibfnamefont {M.~E.}\ \bibnamefont {Tuckerman}}, \bibinfo {author}
  {\bibfnamefont {K.}~\bibnamefont {Burke}}, \ and\ \bibinfo {author}
  {\bibfnamefont {K.~R.}\ \bibnamefont {M{\"u}ller}},\ }\href {\doibase
  10.1038/s41467-017-00839-3} {\bibfield  {journal} {\bibinfo  {journal} {Nat.
  Commun.}\ }\textbf {\bibinfo {volume} {8}},\ \bibinfo {pages} {872} (\bibinfo
  {year} {2017})}\BibitemShut {NoStop}%
\bibitem [{\citenamefont {Alred}\ \emph {et~al.}(2018)\citenamefont {Alred},
  \citenamefont {Bets}, \citenamefont {Xie},\ and\ \citenamefont
  {Yakobson}}]{alre+18cst}%
  \BibitemOpen
  \bibfield  {author} {\bibinfo {author} {\bibfnamefont {J.~M.}\ \bibnamefont
  {Alred}}, \bibinfo {author} {\bibfnamefont {K.~V.}\ \bibnamefont {Bets}},
  \bibinfo {author} {\bibfnamefont {Y.}~\bibnamefont {Xie}}, \ and\ \bibinfo
  {author} {\bibfnamefont {B.~I.}\ \bibnamefont {Yakobson}},\ }\href {\doibase
  10.1016/j.compscitech.2018.03.035} {\bibfield  {journal} {\bibinfo  {journal}
  {Composites Science and Technology}\ }\textbf {\bibinfo {volume} {166}},\
  \bibinfo {pages} {3} (\bibinfo {year} {2018})}\BibitemShut {NoStop}%
\bibitem [{\citenamefont {Chandrasekaran}\ \emph {et~al.}(2019)\citenamefont
  {Chandrasekaran}, \citenamefont {Kamal}, \citenamefont {Batra}, \citenamefont
  {Kim}, \citenamefont {Chen},\ and\ \citenamefont {Ramprasad}}]{chan+19npjcm}%
  \BibitemOpen
  \bibfield  {author} {\bibinfo {author} {\bibfnamefont {A.}~\bibnamefont
  {Chandrasekaran}}, \bibinfo {author} {\bibfnamefont {D.}~\bibnamefont
  {Kamal}}, \bibinfo {author} {\bibfnamefont {R.}~\bibnamefont {Batra}},
  \bibinfo {author} {\bibfnamefont {C.}~\bibnamefont {Kim}}, \bibinfo {author}
  {\bibfnamefont {L.}~\bibnamefont {Chen}}, \ and\ \bibinfo {author}
  {\bibfnamefont {R.}~\bibnamefont {Ramprasad}},\ }\href {\doibase
  10.1038/s41524-019-0162-7} {\bibfield  {journal} {\bibinfo  {journal} {npj
  Comput Mater}\ }\textbf {\bibinfo {volume} {5}},\ \bibinfo {pages} {22}
  (\bibinfo {year} {2019})}\BibitemShut {NoStop}%
\bibitem [{\citenamefont {Grisafi}\ \emph {et~al.}(2019)\citenamefont
  {Grisafi}, \citenamefont {Fabrizio}, \citenamefont {Meyer}, \citenamefont
  {Wilkins}, \citenamefont {Corminboeuf},\ and\ \citenamefont
  {Ceriotti}}]{gris+19acscs}%
  \BibitemOpen
  \bibfield  {author} {\bibinfo {author} {\bibfnamefont {A.}~\bibnamefont
  {Grisafi}}, \bibinfo {author} {\bibfnamefont {A.}~\bibnamefont {Fabrizio}},
  \bibinfo {author} {\bibfnamefont {B.}~\bibnamefont {Meyer}}, \bibinfo
  {author} {\bibfnamefont {D.~M.}\ \bibnamefont {Wilkins}}, \bibinfo {author}
  {\bibfnamefont {C.}~\bibnamefont {Corminboeuf}}, \ and\ \bibinfo {author}
  {\bibfnamefont {M.}~\bibnamefont {Ceriotti}},\ }\href {\doibase
  10.1021/acscentsci.8b00551} {\bibfield  {journal} {\bibinfo  {journal} {ACS
  Cent. Sci.}\ }\textbf {\bibinfo {volume} {5}},\ \bibinfo {pages} {57}
  (\bibinfo {year} {2019})}\BibitemShut {NoStop}%
\bibitem [{\citenamefont {Fabrizio}\ \emph {et~al.}(2019)\citenamefont
  {Fabrizio}, \citenamefont {Grisafi}, \citenamefont {Meyer}, \citenamefont
  {Ceriotti},\ and\ \citenamefont {Corminboeuf}}]{fabr+19cs}%
  \BibitemOpen
  \bibfield  {author} {\bibinfo {author} {\bibfnamefont {A.}~\bibnamefont
  {Fabrizio}}, \bibinfo {author} {\bibfnamefont {A.}~\bibnamefont {Grisafi}},
  \bibinfo {author} {\bibfnamefont {B.}~\bibnamefont {Meyer}}, \bibinfo
  {author} {\bibfnamefont {M.}~\bibnamefont {Ceriotti}}, \ and\ \bibinfo
  {author} {\bibfnamefont {C.}~\bibnamefont {Corminboeuf}},\ }\href {\doibase
  10.1039/C9SC02696G} {\bibfield  {journal} {\bibinfo  {journal} {Chem. Sci.}\
  }\textbf {\bibinfo {volume} {10}},\ \bibinfo {pages} {9424} (\bibinfo {year}
  {2019})}\BibitemShut {NoStop}%
\bibitem [{\citenamefont {Bogojeski}\ \emph {et~al.}(2020)\citenamefont
  {Bogojeski}, \citenamefont {Vogt-Maranto}, \citenamefont {Tuckerman},
  \citenamefont {Müller},\ and\ \citenamefont {Burke}}]{bogo+20nc}%
  \BibitemOpen
  \bibfield  {author} {\bibinfo {author} {\bibfnamefont {M.}~\bibnamefont
  {Bogojeski}}, \bibinfo {author} {\bibfnamefont {L.}~\bibnamefont
  {Vogt-Maranto}}, \bibinfo {author} {\bibfnamefont {M.~E.}\ \bibnamefont
  {Tuckerman}}, \bibinfo {author} {\bibfnamefont {K.-R.}\ \bibnamefont
  {Müller}}, \ and\ \bibinfo {author} {\bibfnamefont {K.}~\bibnamefont
  {Burke}},\ }\href {\doibase 10.1038/s41467-020-19093-1} {\bibfield  {journal}
  {\bibinfo  {journal} {Nat. Commun.}\ }\textbf {\bibinfo {volume} {11}},\
  \bibinfo {pages} {5223} (\bibinfo {year} {2020})}\BibitemShut {NoStop}%
\bibitem [{\citenamefont {J{\o}rgensen}\ and\ \citenamefont
  {Bhowmik}(2022)}]{Jorgensen2021}%
  \BibitemOpen
  \bibfield  {author} {\bibinfo {author} {\bibfnamefont {P.~B.}\ \bibnamefont
  {J{\o}rgensen}}\ and\ \bibinfo {author} {\bibfnamefont {A.}~\bibnamefont
  {Bhowmik}},\ }\href {https://doi.org/10.1038/s41524-022-00863-y} {\bibfield
  {journal} {\bibinfo  {journal} {npj Computational Materials}\ }\textbf
  {\bibinfo {volume} {8}},\ \bibinfo {pages} {183} (\bibinfo {year}
  {2022})}\BibitemShut {NoStop}%
\bibitem [{\citenamefont {Lewis}\ \emph {et~al.}(2021)\citenamefont {Lewis},
  \citenamefont {Grisafi}, \citenamefont {Ceriotti},\ and\ \citenamefont
  {Rossi}}]{lewis+21jctc}%
  \BibitemOpen
  \bibfield  {author} {\bibinfo {author} {\bibfnamefont {A.~M.}\ \bibnamefont
  {Lewis}}, \bibinfo {author} {\bibfnamefont {A.}~\bibnamefont {Grisafi}},
  \bibinfo {author} {\bibfnamefont {M.}~\bibnamefont {Ceriotti}}, \ and\
  \bibinfo {author} {\bibfnamefont {M.}~\bibnamefont {Rossi}},\ }\href
  {\doibase 10.1021/acs.jctc.1c00576} {\bibfield  {journal} {\bibinfo
  {journal} {Journal of Chemical Theory and Computation}\ }\textbf {\bibinfo
  {volume} {17}},\ \bibinfo {pages} {7203} (\bibinfo {year}
  {2021})}\BibitemShut {NoStop}%
\bibitem [{\citenamefont {Grisafi}\ \emph {et~al.}(2023)\citenamefont
  {Grisafi}, \citenamefont {Lewis}, \citenamefont {Rossi},\ and\ \citenamefont
  {Ceriotti}}]{Grisafi2023}%
  \BibitemOpen
  \bibfield  {author} {\bibinfo {author} {\bibfnamefont {A.}~\bibnamefont
  {Grisafi}}, \bibinfo {author} {\bibfnamefont {A.~M.}\ \bibnamefont {Lewis}},
  \bibinfo {author} {\bibfnamefont {M.}~\bibnamefont {Rossi}}, \ and\ \bibinfo
  {author} {\bibfnamefont {M.}~\bibnamefont {Ceriotti}},\ }\href {\doibase
  10.1021/acs.jctc.2c00850} {\bibfield  {journal} {\bibinfo  {journal} {Journal
  of Chemical Theory and Computation}\ }\textbf {\bibinfo {volume} {19}},\
  \bibinfo {pages} {4451} (\bibinfo {year} {2023})}\BibitemShut {NoStop}%
\bibitem [{\citenamefont {Rackers}\ \emph {et~al.}(2023)\citenamefont
  {Rackers}, \citenamefont {Tecot}, \citenamefont {Geiger},\ and\ \citenamefont
  {Smidt}}]{Rackers2023}%
  \BibitemOpen
  \bibfield  {author} {\bibinfo {author} {\bibfnamefont {J.~A.}\ \bibnamefont
  {Rackers}}, \bibinfo {author} {\bibfnamefont {L.}~\bibnamefont {Tecot}},
  \bibinfo {author} {\bibfnamefont {M.}~\bibnamefont {Geiger}}, \ and\ \bibinfo
  {author} {\bibfnamefont {T.~E.}\ \bibnamefont {Smidt}},\ }\href {\doibase
  10.1088/2632-2153/acb314} {\bibfield  {journal} {\bibinfo  {journal} {Machine
  Learning: Science and Technology}\ }\textbf {\bibinfo {volume} {4}},\
  \bibinfo {pages} {015027} (\bibinfo {year} {2023})}\BibitemShut {NoStop}%
\bibitem [{\citenamefont {Grisafi}\ and\ \citenamefont
  {Ceriotti}(2019{\natexlab{a}})}]{Grisafi2019jcp}%
  \BibitemOpen
  \bibfield  {author} {\bibinfo {author} {\bibfnamefont {A.}~\bibnamefont
  {Grisafi}}\ and\ \bibinfo {author} {\bibfnamefont {M.}~\bibnamefont
  {Ceriotti}},\ }\href {https://doi.org/10.1063/1.5128375} {\bibfield
  {journal} {\bibinfo  {journal} {The Journal of Chemical Physics}\ }\textbf
  {\bibinfo {volume} {151}},\ \bibinfo {pages} {204105} (\bibinfo {year}
  {2019}{\natexlab{a}})}\BibitemShut {NoStop}%
\bibitem [{\citenamefont {Gao}\ and\ \citenamefont {Remsing}(2022)}]{Gao2022}%
  \BibitemOpen
  \bibfield  {author} {\bibinfo {author} {\bibfnamefont {A.}~\bibnamefont
  {Gao}}\ and\ \bibinfo {author} {\bibfnamefont {R.~C.}\ \bibnamefont
  {Remsing}},\ }\href {\doibase 10.1038/s41467-022-29243-2} {\bibfield
  {journal} {\bibinfo  {journal} {Nature Communications}\ }\textbf {\bibinfo
  {volume} {13}},\ \bibinfo {pages} {1572} (\bibinfo {year}
  {2022})}\BibitemShut {NoStop}%
\bibitem [{\citenamefont {Zhang}\ \emph {et~al.}(2022)\citenamefont {Zhang},
  \citenamefont {Wang}, \citenamefont {Muniz}, \citenamefont {Panagiotopoulos},
  \citenamefont {Car},\ and\ \citenamefont {E}}]{Zhang-Weinan2022}%
  \BibitemOpen
  \bibfield  {author} {\bibinfo {author} {\bibfnamefont {L.}~\bibnamefont
  {Zhang}}, \bibinfo {author} {\bibfnamefont {H.}~\bibnamefont {Wang}},
  \bibinfo {author} {\bibfnamefont {M.~C.}\ \bibnamefont {Muniz}}, \bibinfo
  {author} {\bibfnamefont {A.~Z.}\ \bibnamefont {Panagiotopoulos}}, \bibinfo
  {author} {\bibfnamefont {R.}~\bibnamefont {Car}}, \ and\ \bibinfo {author}
  {\bibfnamefont {W.}~\bibnamefont {E}},\ }\href {\doibase 10.1063/5.0083669}
  {\bibfield  {journal} {\bibinfo  {journal} {The Journal of Chemical Physics}\
  }\textbf {\bibinfo {volume} {156}},\ \bibinfo {pages} {124107} (\bibinfo
  {year} {2022})}\BibitemShut {NoStop}%
\bibitem [{\citenamefont {Westermayr}\ \emph {et~al.}(2022)\citenamefont
  {Westermayr}, \citenamefont {Chaudhuri}, \citenamefont {Jeindl},
  \citenamefont {Hofmann},\ and\ \citenamefont {Maurer}}]{Westermayr2022}%
  \BibitemOpen
  \bibfield  {author} {\bibinfo {author} {\bibfnamefont {J.}~\bibnamefont
  {Westermayr}}, \bibinfo {author} {\bibfnamefont {S.}~\bibnamefont
  {Chaudhuri}}, \bibinfo {author} {\bibfnamefont {A.}~\bibnamefont {Jeindl}},
  \bibinfo {author} {\bibfnamefont {O.~T.}\ \bibnamefont {Hofmann}}, \ and\
  \bibinfo {author} {\bibfnamefont {R.~J.}\ \bibnamefont {Maurer}},\ }\href
  {\doibase 10.1039/D2DD00016D} {\bibfield  {journal} {\bibinfo  {journal}
  {Digital Discovery}\ }\textbf {\bibinfo {volume} {1}},\ \bibinfo {pages}
  {463} (\bibinfo {year} {2022})}\BibitemShut {NoStop}%
\bibitem [{\citenamefont {Ghasemi}\ \emph {et~al.}(2015)\citenamefont
  {Ghasemi}, \citenamefont {Hofstetter}, \citenamefont {Saha},\ and\
  \citenamefont {Goedecker}}]{Ghasemi2015}%
  \BibitemOpen
  \bibfield  {author} {\bibinfo {author} {\bibfnamefont {S.~A.}\ \bibnamefont
  {Ghasemi}}, \bibinfo {author} {\bibfnamefont {A.}~\bibnamefont {Hofstetter}},
  \bibinfo {author} {\bibfnamefont {S.}~\bibnamefont {Saha}}, \ and\ \bibinfo
  {author} {\bibfnamefont {S.}~\bibnamefont {Goedecker}},\ }\href {\doibase
  10.1103/PhysRevB.92.045131} {\bibfield  {journal} {\bibinfo  {journal} {Phys.
  Rev. B}\ }\textbf {\bibinfo {volume} {92}},\ \bibinfo {pages} {045131}
  (\bibinfo {year} {2015})}\BibitemShut {NoStop}%
\bibitem [{\citenamefont {Ko}\ \emph {et~al.}(2021)\citenamefont {Ko},
  \citenamefont {Finkler}, \citenamefont {Goedecker},\ and\ \citenamefont
  {Behler}}]{Ko2021}%
  \BibitemOpen
  \bibfield  {author} {\bibinfo {author} {\bibfnamefont {T.~W.}\ \bibnamefont
  {Ko}}, \bibinfo {author} {\bibfnamefont {J.~A.}\ \bibnamefont {Finkler}},
  \bibinfo {author} {\bibfnamefont {S.}~\bibnamefont {Goedecker}}, \ and\
  \bibinfo {author} {\bibfnamefont {J.}~\bibnamefont {Behler}},\ }\href
  {\doibase 10.1038/s41467-020-20427-2} {\bibfield  {journal} {\bibinfo
  {journal} {Nature Communications}\ }\textbf {\bibinfo {volume} {12}},\
  \bibinfo {pages} {398} (\bibinfo {year} {2021})}\BibitemShut {NoStop}%
\bibitem [{\citenamefont {Staacke}\ \emph {et~al.}(2022)\citenamefont
  {Staacke}, \citenamefont {Wengert}, \citenamefont {Kunkel}, \citenamefont
  {Cs\'anyi}, \citenamefont {Reuter},\ and\ \citenamefont
  {Margraf}}]{Staacke2022}%
  \BibitemOpen
  \bibfield  {author} {\bibinfo {author} {\bibfnamefont {C.~G.}\ \bibnamefont
  {Staacke}}, \bibinfo {author} {\bibfnamefont {S.}~\bibnamefont {Wengert}},
  \bibinfo {author} {\bibfnamefont {C.}~\bibnamefont {Kunkel}}, \bibinfo
  {author} {\bibfnamefont {G.}~\bibnamefont {Cs\'anyi}}, \bibinfo {author}
  {\bibfnamefont {K.}~\bibnamefont {Reuter}}, \ and\ \bibinfo {author}
  {\bibfnamefont {J.~T.}\ \bibnamefont {Margraf}},\ }\href {\doibase
  10.1088/2632-2153/ac568d} {\bibfield  {journal} {\bibinfo  {journal} {Machine
  Learning: Science and Technology}\ }\textbf {\bibinfo {volume} {3}},\
  \bibinfo {pages} {015032} (\bibinfo {year} {2022})}\BibitemShut {NoStop}%
\bibitem [{\citenamefont {Shao}\ \emph {et~al.}(2022)\citenamefont {Shao},
  \citenamefont {Andersson}, \citenamefont {Knijff},\ and\ \citenamefont
  {Zhang}}]{Shao2022}%
  \BibitemOpen
  \bibfield  {author} {\bibinfo {author} {\bibfnamefont {Y.}~\bibnamefont
  {Shao}}, \bibinfo {author} {\bibfnamefont {L.}~\bibnamefont {Andersson}},
  \bibinfo {author} {\bibfnamefont {L.}~\bibnamefont {Knijff}}, \ and\ \bibinfo
  {author} {\bibfnamefont {C.}~\bibnamefont {Zhang}},\ }\href {\doibase
  10.1088/2516-1075/ac59ca} {\bibfield  {journal} {\bibinfo  {journal}
  {Electronic Structure}\ }\textbf {\bibinfo {volume} {4}},\ \bibinfo {pages}
  {014012} (\bibinfo {year} {2022})}\BibitemShut {NoStop}%
\bibitem [{\citenamefont {Dufils}\ \emph {et~al.}(2023)\citenamefont {Dufils},
  \citenamefont {Knijff}, \citenamefont {Shao},\ and\ \citenamefont
  {Zhang}}]{Dufils2023arxiv}%
  \BibitemOpen
  \bibfield  {author} {\bibinfo {author} {\bibfnamefont {T.}~\bibnamefont
  {Dufils}}, \bibinfo {author} {\bibfnamefont {L.}~\bibnamefont {Knijff}},
  \bibinfo {author} {\bibfnamefont {Y.}~\bibnamefont {Shao}}, \ and\ \bibinfo
  {author} {\bibfnamefont {C.}~\bibnamefont {Zhang}},\ }\href
  {http://arxiv.org/abs/2303.15307} {\bibfield  {journal} {\bibinfo  {journal}
  {ArXiv Prepr. ArXiv2303.15307}\ } (\bibinfo {year} {2023})}\BibitemShut
  {NoStop}%
\bibitem [{\citenamefont {Grisafi}\ \emph {et~al.}(2018)\citenamefont
  {Grisafi}, \citenamefont {Wilkins}, \citenamefont {Cs{\'a}nyi},\ and\
  \citenamefont {Ceriotti}}]{gris+18prl}%
  \BibitemOpen
  \bibfield  {author} {\bibinfo {author} {\bibfnamefont {A.}~\bibnamefont
  {Grisafi}}, \bibinfo {author} {\bibfnamefont {D.~M.}\ \bibnamefont
  {Wilkins}}, \bibinfo {author} {\bibfnamefont {G.}~\bibnamefont {Cs{\'a}nyi}},
  \ and\ \bibinfo {author} {\bibfnamefont {M.}~\bibnamefont {Ceriotti}},\
  }\href {\doibase 10.1103/PhysRevLett.120.036002} {\bibfield  {journal}
  {\bibinfo  {journal} {Phys. Rev. Lett.}\ }\textbf {\bibinfo {volume} {120}},\
  \bibinfo {pages} {036002} (\bibinfo {year} {2018})}\BibitemShut {NoStop}%
\bibitem [{\citenamefont {Musil}\ \emph {et~al.}(2021)\citenamefont {Musil},
  \citenamefont {Grisafi}, \citenamefont {Bart\'ok}, \citenamefont {Ortner},
  \citenamefont {Cs\'anyi},\ and\ \citenamefont {Ceriotti}}]{musil_chem_rev}%
  \BibitemOpen
  \bibfield  {author} {\bibinfo {author} {\bibfnamefont {F.}~\bibnamefont
  {Musil}}, \bibinfo {author} {\bibfnamefont {A.}~\bibnamefont {Grisafi}},
  \bibinfo {author} {\bibfnamefont {A.~P.}\ \bibnamefont {Bart\'ok}}, \bibinfo
  {author} {\bibfnamefont {C.}~\bibnamefont {Ortner}}, \bibinfo {author}
  {\bibfnamefont {G.}~\bibnamefont {Cs\'anyi}}, \ and\ \bibinfo {author}
  {\bibfnamefont {M.}~\bibnamefont {Ceriotti}},\ }\href {\doibase
  10.1021/acs.chemrev.1c00021} {\bibfield  {journal} {\bibinfo  {journal}
  {Chemical Reviews}\ }\textbf {\bibinfo {volume} {121}},\ \bibinfo {pages}
  {9759} (\bibinfo {year} {2021})}\BibitemShut {NoStop}%
\bibitem [{\citenamefont {Deringer}\ \emph {et~al.}(2021)\citenamefont
  {Deringer}, \citenamefont {Bart\'ok}, \citenamefont {Bernstein},
  \citenamefont {Wilkins}, \citenamefont {Ceriotti},\ and\ \citenamefont
  {Cs\'anyi}}]{deringer_chem_rev}%
  \BibitemOpen
  \bibfield  {author} {\bibinfo {author} {\bibfnamefont {V.~L.}\ \bibnamefont
  {Deringer}}, \bibinfo {author} {\bibfnamefont {A.~P.}\ \bibnamefont
  {Bart\'ok}}, \bibinfo {author} {\bibfnamefont {N.}~\bibnamefont {Bernstein}},
  \bibinfo {author} {\bibfnamefont {D.~M.}\ \bibnamefont {Wilkins}}, \bibinfo
  {author} {\bibfnamefont {M.}~\bibnamefont {Ceriotti}}, \ and\ \bibinfo
  {author} {\bibfnamefont {G.}~\bibnamefont {Cs\'anyi}},\ }\href {\doibase
  10.1021/acs.chemrev.1c00022} {\bibfield  {journal} {\bibinfo  {journal}
  {Chemical Reviews}\ }\textbf {\bibinfo {volume} {121}},\ \bibinfo {pages}
  {10073} (\bibinfo {year} {2021})}\BibitemShut {NoStop}%
\bibitem [{\citenamefont {Bart{\'o}k}\ \emph {et~al.}(2013)\citenamefont
  {Bart{\'o}k}, \citenamefont {Kondor},\ and\ \citenamefont
  {Cs{\'a}nyi}}]{bart+13prb}%
  \BibitemOpen
  \bibfield  {author} {\bibinfo {author} {\bibfnamefont {A.~P.}\ \bibnamefont
  {Bart{\'o}k}}, \bibinfo {author} {\bibfnamefont {R.}~\bibnamefont {Kondor}},
  \ and\ \bibinfo {author} {\bibfnamefont {G.}~\bibnamefont {Cs{\'a}nyi}},\
  }\href {\doibase 10.1103/PhysRevB.87.184115} {\bibfield  {journal} {\bibinfo
  {journal} {Phys. Rev. B}\ }\textbf {\bibinfo {volume} {87}},\ \bibinfo
  {pages} {184115} (\bibinfo {year} {2013})}\BibitemShut {NoStop}%
\bibitem [{\citenamefont {Fabrizio}\ \emph {et~al.}(2020)\citenamefont
  {Fabrizio}, \citenamefont {Briling}, \citenamefont {Grisafi},\ and\
  \citenamefont {Corminboeuf}}]{fabr+20chimia}%
  \BibitemOpen
  \bibfield  {author} {\bibinfo {author} {\bibfnamefont {A.}~\bibnamefont
  {Fabrizio}}, \bibinfo {author} {\bibfnamefont {K.}~\bibnamefont {Briling}},
  \bibinfo {author} {\bibfnamefont {A.}~\bibnamefont {Grisafi}}, \ and\
  \bibinfo {author} {\bibfnamefont {C.}~\bibnamefont {Corminboeuf}},\ }\href
  {\doibase doi:10.2533/chimia.2020.232} {\bibfield  {journal} {\bibinfo
  {journal} {CHIMIA International Journal for Chemistry}\ }\textbf {\bibinfo
  {volume} {74}},\ \bibinfo {pages} {232} (\bibinfo {year} {2020})}\BibitemShut
  {NoStop}%
\bibitem [{\citenamefont {Grisafi}\ and\ \citenamefont
  {Ceriotti}(2019{\natexlab{b}})}]{gris-ceri19jcp}%
  \BibitemOpen
  \bibfield  {author} {\bibinfo {author} {\bibfnamefont {A.}~\bibnamefont
  {Grisafi}}\ and\ \bibinfo {author} {\bibfnamefont {M.}~\bibnamefont
  {Ceriotti}},\ }\href {\doibase 10.1063/1.5128375} {\bibfield  {journal}
  {\bibinfo  {journal} {J. Chem. Phys.}\ }\textbf {\bibinfo {volume} {151}},\
  \bibinfo {pages} {204105} (\bibinfo {year} {2019}{\natexlab{b}})}\BibitemShut
  {NoStop}%
\bibitem [{\citenamefont {Grisafi}\ \emph {et~al.}(2021)\citenamefont
  {Grisafi}, \citenamefont {Nigam},\ and\ \citenamefont
  {Ceriotti}}]{grisafi2021cs}%
  \BibitemOpen
  \bibfield  {author} {\bibinfo {author} {\bibfnamefont {A.}~\bibnamefont
  {Grisafi}}, \bibinfo {author} {\bibfnamefont {J.}~\bibnamefont {Nigam}}, \
  and\ \bibinfo {author} {\bibfnamefont {M.}~\bibnamefont {Ceriotti}},\ }\href
  {\doibase 10.1039/D0SC04934D} {\bibfield  {journal} {\bibinfo  {journal}
  {Chem. Sci.}\ }\textbf {\bibinfo {volume} {12}},\ \bibinfo {pages} {2078}
  (\bibinfo {year} {2021})}\BibitemShut {NoStop}%
\bibitem [{\citenamefont {Willatt}\ \emph {et~al.}(2019)\citenamefont
  {Willatt}, \citenamefont {Musil},\ and\ \citenamefont
  {Ceriotti}}]{willatt2019}%
  \BibitemOpen
  \bibfield  {author} {\bibinfo {author} {\bibfnamefont {M.~J.}\ \bibnamefont
  {Willatt}}, \bibinfo {author} {\bibfnamefont {F.}~\bibnamefont {Musil}}, \
  and\ \bibinfo {author} {\bibfnamefont {M.}~\bibnamefont {Ceriotti}},\ }\href
  {https://doi.org/10.1063/1.5090481} {\bibfield  {journal} {\bibinfo
  {journal} {The Journal of Chemical Physics}\ }\textbf {\bibinfo {volume}
  {150}},\ \bibinfo {pages} {154110} (\bibinfo {year} {2019})}\BibitemShut
  {NoStop}%
\bibitem [{\citenamefont {Laino}\ \emph {et~al.}(2006)\citenamefont {Laino},
  \citenamefont {Mohamed}, \citenamefont {Laio},\ and\ \citenamefont
  {Parrinello}}]{Laino2006}%
  \BibitemOpen
  \bibfield  {author} {\bibinfo {author} {\bibfnamefont {T.}~\bibnamefont
  {Laino}}, \bibinfo {author} {\bibfnamefont {F.}~\bibnamefont {Mohamed}},
  \bibinfo {author} {\bibfnamefont {A.}~\bibnamefont {Laio}}, \ and\ \bibinfo
  {author} {\bibfnamefont {M.}~\bibnamefont {Parrinello}},\ }\href {\doibase
  10.1021/ct6001169} {\bibfield  {journal} {\bibinfo  {journal} {Journal of
  Chemical Theory and Computation}\ }\textbf {\bibinfo {volume} {2}},\ \bibinfo
  {pages} {1370} (\bibinfo {year} {2006})}\BibitemShut {NoStop}%
\bibitem [{\citenamefont {Perdew}\ \emph {et~al.}(1996)\citenamefont {Perdew},
  \citenamefont {Burke},\ and\ \citenamefont {Ernzerhof}}]{Perdew1996}%
  \BibitemOpen
  \bibfield  {author} {\bibinfo {author} {\bibfnamefont {J.~P.}\ \bibnamefont
  {Perdew}}, \bibinfo {author} {\bibfnamefont {K.}~\bibnamefont {Burke}}, \
  and\ \bibinfo {author} {\bibfnamefont {M.}~\bibnamefont {Ernzerhof}},\ }\href
  {\doibase 10.1103/PhysRevLett.77.3865} {\bibfield  {journal} {\bibinfo
  {journal} {Phys. Rev. Lett.}\ }\textbf {\bibinfo {volume} {77}},\ \bibinfo
  {pages} {3865} (\bibinfo {year} {1996})}\BibitemShut {NoStop}%
\bibitem [{\citenamefont {Vahtras}\ \emph {et~al.}(1993)\citenamefont
  {Vahtras}, \citenamefont {Almlöf},\ and\ \citenamefont
  {Feyereisen}}]{Vahtras1993}%
  \BibitemOpen
  \bibfield  {author} {\bibinfo {author} {\bibfnamefont {O.}~\bibnamefont
  {Vahtras}}, \bibinfo {author} {\bibfnamefont {J.}~\bibnamefont {Almlöf}}, \
  and\ \bibinfo {author} {\bibfnamefont {M.}~\bibnamefont {Feyereisen}},\
  }\href {\doibase https://doi.org/10.1016/0009-2614(93)89151-7} {\bibfield
  {journal} {\bibinfo  {journal} {Chemical Physics Letters}\ }\textbf {\bibinfo
  {volume} {213}},\ \bibinfo {pages} {514} (\bibinfo {year}
  {1993})}\BibitemShut {NoStop}%
\bibitem [{\citenamefont {Briling}\ \emph {et~al.}(2021)\citenamefont
  {Briling}, \citenamefont {Fabrizio},\ and\ \citenamefont
  {Corminboeuf}}]{briling2021}%
  \BibitemOpen
  \bibfield  {author} {\bibinfo {author} {\bibfnamefont {K.~R.}\ \bibnamefont
  {Briling}}, \bibinfo {author} {\bibfnamefont {A.}~\bibnamefont {Fabrizio}}, \
  and\ \bibinfo {author} {\bibfnamefont {C.}~\bibnamefont {Corminboeuf}},\
  }\href {https://doi.org/10.1063/5.0055393} {\bibfield  {journal} {\bibinfo
  {journal} {The Journal of Chemical Physics}\ }\textbf {\bibinfo {volume}
  {155}},\ \bibinfo {pages} {024107} (\bibinfo {year} {2021})}\BibitemShut
  {NoStop}%
\bibitem [{\citenamefont {Bussy}\ \emph {et~al.}(2023)\citenamefont {Bussy},
  \citenamefont {Schütt},\ and\ \citenamefont {Hutter}}]{Bussy2023}%
  \BibitemOpen
  \bibfield  {author} {\bibinfo {author} {\bibfnamefont {A.}~\bibnamefont
  {Bussy}}, \bibinfo {author} {\bibfnamefont {O.}~\bibnamefont {Schütt}}, \
  and\ \bibinfo {author} {\bibfnamefont {J.}~\bibnamefont {Hutter}},\ }\href
  {\doibase 10.1063/5.0144493} {\bibfield  {journal} {\bibinfo  {journal} {The
  Journal of Chemical Physics}\ }\textbf {\bibinfo {volume} {158}},\ \bibinfo
  {pages} {164109} (\bibinfo {year} {2023})}\BibitemShut {NoStop}%
\bibitem [{\citenamefont {Willatt}\ \emph {et~al.}(2018)\citenamefont
  {Willatt}, \citenamefont {Musil},\ and\ \citenamefont
  {Ceriotti}}]{will+18pccp}%
  \BibitemOpen
  \bibfield  {author} {\bibinfo {author} {\bibfnamefont {M.~J.}\ \bibnamefont
  {Willatt}}, \bibinfo {author} {\bibfnamefont {F.}~\bibnamefont {Musil}}, \
  and\ \bibinfo {author} {\bibfnamefont {M.}~\bibnamefont {Ceriotti}},\ }\href
  {\doibase 10.1039/c8cp05921g} {\bibfield  {journal} {\bibinfo  {journal}
  {Phys. Chem. Chem. Phys.}\ }\textbf {\bibinfo {volume} {20}},\ \bibinfo
  {pages} {29661} (\bibinfo {year} {2018})}\BibitemShut {NoStop}%
\bibitem [{sup()}]{suppmat}%
  \BibitemOpen
  \href@noop {} {}\bibinfo {note} {See Supplemental Material at \textit{[URL
  will be set by the Editor]}, which contains: details about the dataset
  generation, a derivation of the finite-field extension of LODE, details about
  the ML parameters, a discussion about charge conservation, complementary
  results, an explicit calculation of the electronic polarization vector, a
  discussion about the calculation the differential capacitance.}\BibitemShut
  {Stop}%
\bibitem [{\citenamefont {Theodoridis}\ and\ \citenamefont
  {Koutroumbas}(2009)}]{Konstantinos2009}%
  \BibitemOpen
  \bibfield  {author} {\bibinfo {author} {\bibfnamefont {S.}~\bibnamefont
  {Theodoridis}}\ and\ \bibinfo {author} {\bibfnamefont {K.}~\bibnamefont
  {Koutroumbas}},\ }\href
  {https://www.bibsonomy.org/bibtex/284b7e7aa970c54eebb504c44f3e29535/procomun}
  {\emph {\bibinfo {title} {{Pattern Recognition, Fourth Edition}}}}\ (\bibinfo
   {publisher} {{Academic Press}},\ \bibinfo {year} {2009})\BibitemShut
  {NoStop}%
\bibitem [{\citenamefont {Dufils}\ \emph {et~al.}(2019)\citenamefont {Dufils},
  \citenamefont {Jeanmairet}, \citenamefont {Rotenberg}, \citenamefont
  {Sprik},\ and\ \citenamefont {Salanne}}]{Dufils2019}%
  \BibitemOpen
  \bibfield  {author} {\bibinfo {author} {\bibfnamefont {T.}~\bibnamefont
  {Dufils}}, \bibinfo {author} {\bibfnamefont {G.}~\bibnamefont {Jeanmairet}},
  \bibinfo {author} {\bibfnamefont {B.}~\bibnamefont {Rotenberg}}, \bibinfo
  {author} {\bibfnamefont {M.}~\bibnamefont {Sprik}}, \ and\ \bibinfo {author}
  {\bibfnamefont {M.}~\bibnamefont {Salanne}},\ }\href {\doibase
  10.1103/PhysRevLett.123.195501} {\bibfield  {journal} {\bibinfo  {journal}
  {Phys. Rev. Lett.}\ }\textbf {\bibinfo {volume} {123}},\ \bibinfo {pages}
  {195501} (\bibinfo {year} {2019})}\BibitemShut {NoStop}%
\bibitem [{\citenamefont {Zhang}\ \emph {et~al.}(2020)\citenamefont {Zhang},
  \citenamefont {Sayer}, \citenamefont {Hutter},\ and\ \citenamefont
  {Sprik}}]{Zhang2020}%
  \BibitemOpen
  \bibfield  {author} {\bibinfo {author} {\bibfnamefont {C.}~\bibnamefont
  {Zhang}}, \bibinfo {author} {\bibfnamefont {T.}~\bibnamefont {Sayer}},
  \bibinfo {author} {\bibfnamefont {H.}~\bibnamefont {Hutter}}, \ and\ \bibinfo
  {author} {\bibfnamefont {M.}~\bibnamefont {Sprik}},\ }\href {\doibase
  10.1088/2515-7655/ab9d8c} {\bibfield  {journal} {\bibinfo  {journal} {Journal
  of Physics: Energy}\ }\textbf {\bibinfo {volume} {2}},\ \bibinfo {pages}
  {032005} (\bibinfo {year} {2020})}\BibitemShut {NoStop}%
\bibitem [{\citenamefont {Marin-Laflèche}\ \emph {et~al.}(2020)\citenamefont
  {Marin-Laflèche}, \citenamefont {Haefele}, \citenamefont {Scalfi},
  \citenamefont {Coretti}, \citenamefont {Dufils}, \citenamefont {Jeanmairet},
  \citenamefont {Reed}, \citenamefont {Serva}, \citenamefont {Berthin},
  \citenamefont {Bacon}, \citenamefont {Bonella}, \citenamefont {Rotenberg},
  \citenamefont {Madden},\ and\ \citenamefont {Salanne}}]{Marin-Lafleche2020}%
  \BibitemOpen
  \bibfield  {author} {\bibinfo {author} {\bibfnamefont {A.}~\bibnamefont
  {Marin-Laflèche}}, \bibinfo {author} {\bibfnamefont {M.}~\bibnamefont
  {Haefele}}, \bibinfo {author} {\bibfnamefont {L.}~\bibnamefont {Scalfi}},
  \bibinfo {author} {\bibfnamefont {A.}~\bibnamefont {Coretti}}, \bibinfo
  {author} {\bibfnamefont {T.}~\bibnamefont {Dufils}}, \bibinfo {author}
  {\bibfnamefont {G.}~\bibnamefont {Jeanmairet}}, \bibinfo {author}
  {\bibfnamefont {S.~K.}\ \bibnamefont {Reed}}, \bibinfo {author}
  {\bibfnamefont {A.}~\bibnamefont {Serva}}, \bibinfo {author} {\bibfnamefont
  {R.}~\bibnamefont {Berthin}}, \bibinfo {author} {\bibfnamefont
  {C.}~\bibnamefont {Bacon}}, \bibinfo {author} {\bibfnamefont
  {S.}~\bibnamefont {Bonella}}, \bibinfo {author} {\bibfnamefont
  {B.}~\bibnamefont {Rotenberg}}, \bibinfo {author} {\bibfnamefont {P.~A.}\
  \bibnamefont {Madden}}, \ and\ \bibinfo {author} {\bibfnamefont
  {M.}~\bibnamefont {Salanne}},\ }\href {\doibase 10.21105/joss.02373}
  {\bibfield  {journal} {\bibinfo  {journal} {Journal of Open Source Software}\
  }\textbf {\bibinfo {volume} {5}},\ \bibinfo {pages} {2373} (\bibinfo {year}
  {2020})}\BibitemShut {NoStop}%
\bibitem [{\citenamefont {Darden}\ \emph {et~al.}(1993)\citenamefont {Darden},
  \citenamefont {York},\ and\ \citenamefont {Pedersen}}]{Darden1993}%
  \BibitemOpen
  \bibfield  {author} {\bibinfo {author} {\bibfnamefont {T.}~\bibnamefont
  {Darden}}, \bibinfo {author} {\bibfnamefont {D.}~\bibnamefont {York}}, \ and\
  \bibinfo {author} {\bibfnamefont {L.}~\bibnamefont {Pedersen}},\ }\href
  {\doibase 10.1063/1.464397} {\bibfield  {journal} {\bibinfo  {journal} {The
  Journal of Chemical Physics}\ }\textbf {\bibinfo {volume} {98}},\ \bibinfo
  {pages} {10089} (\bibinfo {year} {1993})}\BibitemShut {NoStop}%
\bibitem [{\citenamefont {Lautar}\ \emph
  {et~al.}(2020{\natexlab{b}})\citenamefont {Lautar}, \citenamefont
  {Hagopian},\ and\ \citenamefont {Filhol}}]{Lautar2020}%
  \BibitemOpen
  \bibfield  {author} {\bibinfo {author} {\bibfnamefont {A.~K.}\ \bibnamefont
  {Lautar}}, \bibinfo {author} {\bibfnamefont {A.}~\bibnamefont {Hagopian}}, \
  and\ \bibinfo {author} {\bibfnamefont {J.-S.}\ \bibnamefont {Filhol}},\
  }\href {https://doi.org/10.1039/C9CP06684E} {\bibfield  {journal} {\bibinfo
  {journal} {Phys. Chem. Chem. Phys.}\ }\textbf {\bibinfo {volume} {22}},\
  \bibinfo {pages} {10569} (\bibinfo {year} {2020}{\natexlab{b}})}\BibitemShut
  {NoStop}%
\bibitem [{\citenamefont {Lewis}\ \emph {et~al.}(2023)\citenamefont {Lewis},
  \citenamefont {Lazzaroni},\ and\ \citenamefont {Rossi}}]{Lewis2023}%
  \BibitemOpen
  \bibfield  {author} {\bibinfo {author} {\bibfnamefont {A.~M.}\ \bibnamefont
  {Lewis}}, \bibinfo {author} {\bibfnamefont {P.}~\bibnamefont {Lazzaroni}}, \
  and\ \bibinfo {author} {\bibfnamefont {M.}~\bibnamefont {Rossi}},\ }\href
  {\doibase 10.1063/5.0154710} {\bibfield  {journal} {\bibinfo  {journal} {The
  Journal of Chemical Physics}\ }\textbf {\bibinfo {volume} {159}},\ \bibinfo
  {pages} {014103} (\bibinfo {year} {2023})}\BibitemShut {NoStop}%
\bibitem [{\citenamefont {Chowdhury}\ \emph {et~al.}(2021)\citenamefont
  {Chowdhury}, \citenamefont {Muralidharan}, \citenamefont {Daniel},
  \citenamefont {Amin},\ and\ \citenamefont {Belharouak}}]{Chowdhury2021}%
  \BibitemOpen
  \bibfield  {author} {\bibinfo {author} {\bibfnamefont {A.~U.}\ \bibnamefont
  {Chowdhury}}, \bibinfo {author} {\bibfnamefont {N.}~\bibnamefont
  {Muralidharan}}, \bibinfo {author} {\bibfnamefont {C.}~\bibnamefont
  {Daniel}}, \bibinfo {author} {\bibfnamefont {R.}~\bibnamefont {Amin}}, \ and\
  \bibinfo {author} {\bibfnamefont {I.}~\bibnamefont {Belharouak}},\ }\href
  {\doibase https://doi.org/10.1016/j.jpowsour.2021.230173} {\bibfield
  {journal} {\bibinfo  {journal} {Journal of Power Sources}\ }\textbf {\bibinfo
  {volume} {506}},\ \bibinfo {pages} {230173} (\bibinfo {year}
  {2021})}\BibitemShut {NoStop}%
\end{thebibliography}
\end{document}